# Boron Nitride and Oxide Supported on Dendritic Fibrous Nanosilica for Catalytic Oxidative Dehydrogenation of Propane

Rajesh Belgamwar,[1] Andrew G. M. Rankin,[2#] Ayan Maity,[1] Amit Kumar Mishra,[1] Jennifer S. Gómez,[2] Julien Trébosc,[2,4] Chathakudath P. Vinod,[3] Olivier Lafon,[2,5*] Vivek Polshettiwar[1*]

[1]Department of Chemical Sciences, Tata Institute of Fundamental Research (TIFR), Mumbai, India.
*Email: vivekpol@tifr.res.in
[2]Univ. Lille, CNRS, Centrale Lille, ENSCL, Univ. Artois, UMR 8181, UCCS – Unité de Catalyse et Chimie du Solide, F-59000 Lille, France. *Email: olivier.lafon@univ-lille.fr
[3] Catalysis and Inorganic Chemistry Division, CSIR-National Chemical Laboratory (NCL), Pune, India
[4]Univ. Lille, CNRS-FR2638, Fédération Chevreul, F-59000 Lille, France
[5]Institut Universitaire de France, 1 rue Descartes, F-75231 Paris Cedex 05, France
[#]Present Address: Sorbonne Université, CNRS, Collège de France, Laboratoire de Chimie de la Matière Condensée de Paris (LCMCP), 4 place Jussieu, F-75005 Paris, France

**Abstract**

In this work, we were able to significantly increase the activity of boron nitride catalysts used for the oxidative dehydrogenation (ODH) of propane by designing and synthesising boron nitride (BN) supported on dendritic fibrous nanosilica (DFNS). DFNS/BN showed a markedly increased catalytic efficiency, accompanied by exceptional stability and selectivity. Textural characterisation together with solid-state NMR and X-ray photoelectron spectroscopic analyses indicate the presence of a combination of unique fibrous morphology of DFNS and various boron sites connected to silica to be the reason for this increase in the catalytic performance. Notably, DFNS/$B_2O_3$ also showed catalytic activity, although with more moderate selectivity compared to that of DFNS/BN. Solid-state NMR spectra indicates that the higher selectivity of DFNS/BN might stem from a larger amount of hydrogen-bonded hydroxyl groups attached to B atoms. This study indicates that both boron nitride and oxide are active catalysts and by using high surface area support (DFNS), conversion from propane to propene as well as productivity of olefins was significantly increased.

## 1. Introduction

Propene is a very important building block for a large number of chemical products, such as the widely used poly(propene) polymer. It is also used to produce a large quantity of other useful chemicals, such as acrolein, acrylonitrile, cumene, propylene oxide and butyraldehyde. The catalytic oxidative dehydrogenation of alkanes to produce alkenes can be a game-changing technology in the chemical industry.[1-4] However, even after decades of research, the propene selectivity remains too low due to over-oxidation of propene to $CO_2$. Therefore, there is an urgent need for a heterogeneous catalytic process which can convert propane to propene at low temperatures and with good selectivity.

Recently catalytic properties of hexagonal boron nitride (*h*-BN) and boron nitride nanotubes (BNNT) for the conversion of propane to propene with high selectivity to olefins was reported.[5-6] This discovery provides an alternative route for the production of olefins from alkanes, which is much more sustainable than the conventional protocol. It was hypothesized that the oxidised edges of these materials were the catalytic active sites.[5] Additional studies on various boron-containing materials with oxidised surfaces supported that assumption.[7-10] In particular, $^{11}$B solid-state NMR and in operando infrared spectroscopies showed the oxidation of *h*-BN surface under ODH conditions.[8,10] DFT calculations also proved that the formation of a disordered boron oxide phase at the surface of *h*-BN is feasible.[11] More recently it has been shown that boron oxide supported on mesoporous silica or silica nanoparticles can catalyze the ODH of propane.[12,13] Conversely it was shown that zeolites containing isolated $BO_3$ units incorporated into the zeolite framework shows no catalytic activity for the ODH of propane to propene and the catalysis of ODH reaction requires the presence of aggregated boron sites.[14] In this work, we have sought to ascertain whether the activity and selectivity of these catalysts can be increased and the structures made more stable.

Herein we report the design and synthesis of boron nitride supported on dendritic fibrous nanosilica (DFNS/BN). We used DFNS as a support because it has large surface area and excellent physical and textural properties, as well as possessing a unique fibrous morphology instead of the porous structures found in, for example, MCM-41 and SBA-15 type materials.[15-17]

## 2. Results and discussion

We coated the DFNS fibres with BN using boric acid and urea as precursors using our recently developed solution phase deposition process,[17] which was then treated at a temperature of 900 °C under nitrogen for 5 h to produce DFNS/BN. The crystallographic forms of BN, oxidation states, and textural properties (surface area, pore volume, pore sizes) were determined by transmission and scanning electron microscopy (TEM and SEM), energy-dispersive X-ray spectroscopy (EDS), powder X-ray diffraction (PXRD), $N_2$ sorption analysis, and X-ray photoelectron spectroscopy (XPS). The local structure of these materials was also studied by $^1$H, $^{11}$B and $^{29}$Si solid-state nuclear magnetic resonance (NMR) spectroscopy.

Figures 1a and b show the SEM and TEM images of DFNS/BN. They reveal that the fibrous morphology of DFNS remained intact and that a uniformly coated BN layer was achieved. No superfluous BN particle formation outside of the DFNS nanospheres was observed. The BN loading on DFNS was estimated by SEM-EDX analysis at 10 different points (to minimise errors), and was found to be 13.7 wt.%. Its nitrogen sorption isotherm is of type IV, with a type H4 hysteresis (Figure 1c), typical of mesoporous materials lacking well-defined pores.[18-22] The BET surface area (SA) and BJH cumulative pore volume (PV) of DFNS/BN were determined to be 33 m$^2$.g$^{-1}$ and 0.06 cm$^3$.g$^{-1}$ respectively, for DFNS/BN. The reduced surface area and pore volume in this material compared to that of the DFNS before coating (609 m$^2$.g$^{-1}$ and 0.8 cm$^3$.g$^{-1}$ respectively), indicate the filling of the DFNS channels by BN, thus resulting in an increased fibre thickness and decreased space between fibres.

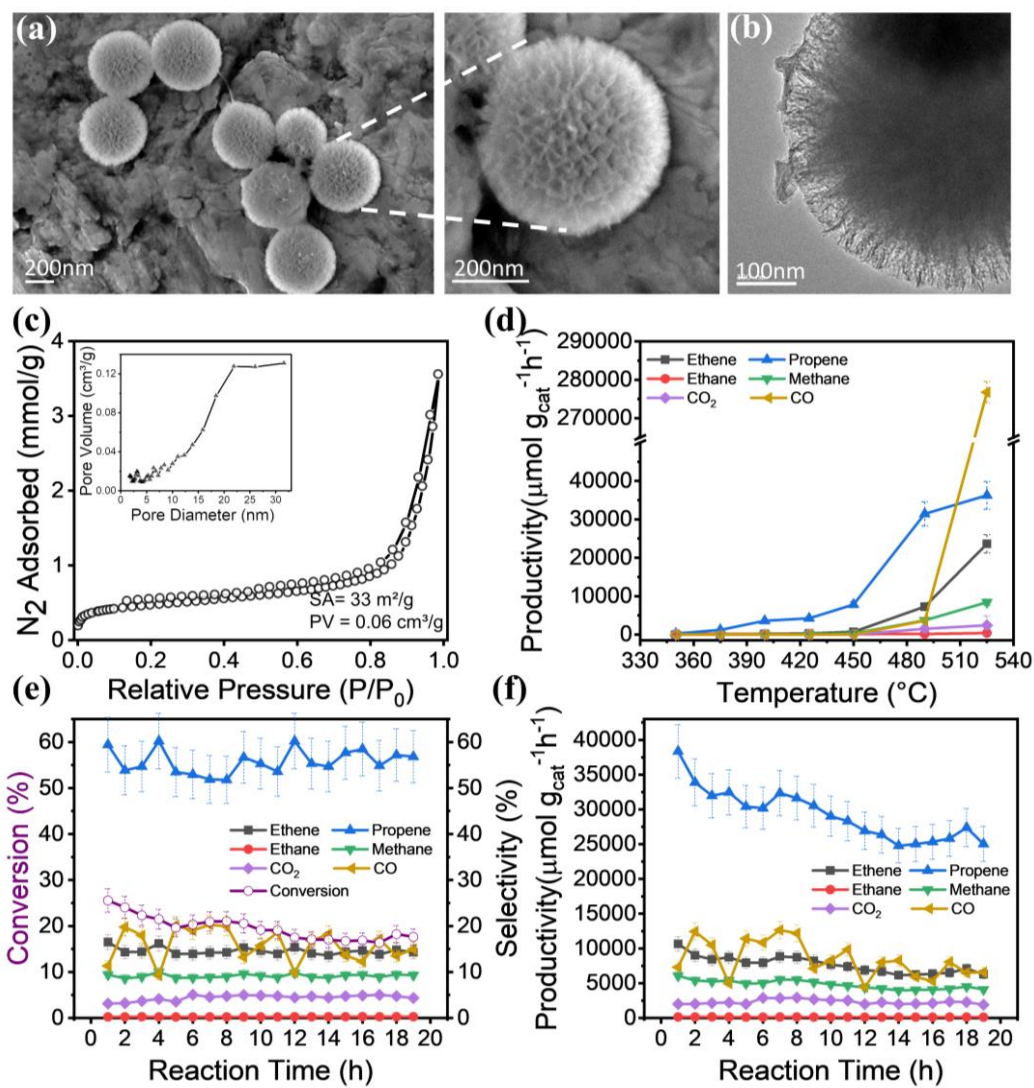

**Figure 1.** (a) SEM, (b) TEM and (c) N$_2$ sorption isotherm (pore size distribution using absorption data is shown in inset) of DFNS/BN catalysts. (d) Effect of temperature on productivity for propane dehydrogenation using DFNS/BN catalysts. (e, f) Time evolution of (e) conversion and selectivity as well as (f) productivity for the same reaction and catalyst as in subfigure (d) carried out at 490 °C. Background in SEM images is a rough surface of the aluminum grid. Error bars were calculated by conducting reactions three times.

The catalytic activity of DFNS/BN for propane dehydrogenation was studied using a fixed bed micro-activity flow reactor under atmospheric pressure (1 bar). We first optimised the reaction temperature (Figure 1d) and 490 °C was found to be the best temperature in terms of catalytic activity and selectivity. At 490 °C, we studied the stability of DFNS/BN on the basis of conversion and selectivity (Figure 1e) and observed that the catalyst was moderately stable for 19 h with decrease in conversion to ~ 18 % with around 59.4% selectivity for propene and around 16.5% for ethene. In terms of productivity, a significant amount of propene (38360 µmol.g$_{cat}^{-1}$.h$^{-1}$) and ethene (10657 µmol.g$_{cat}^{-1}$.h$^{-1}$) was produced. However, the productivity of propene decreased with time (Figure 1f) indicating coke formation (Figure S1) on the catalyst surface.

In order to gain insight into the mechanistic details of this propane conversion, we performed PXRD, and XPS studies of DFNS/BN before and after catalysis (Figure 2). PXRD data indicate that as-synthesized DFNS/BN contains a crystalline BN phase (JCPDS-00-018-0251) (Figure 2a). After catalysis, the diffraction peaks of crystalline BN phase disappear, whereas a boron oxide phase (JCPDS-00-013-0570) is detected. XPS studies also evidence the conversion of BN phase to boron oxide phase after catalysis. Figure 2b-e shows typical B1s, Si2p, O1s and N1s core-level spectra. The B1s spectrum of as-prepared DFNS/BN (Figure 2b) is the superposition of peaks at a binding energy of 190 and 192-193 eV assigned to B-N supported on silica and O-B-O, B-O-H environments, respectively. It indicates that before catalysis, DFNS/BN contains both phases, boron nitride and boron oxide. Since boron oxide was not observed in PXRD, it must be in the amorphous phase. After catalysis, the B-N signal at 190 eV disappears after catalysis, while the broad signal around 192-193 eV assigned to boron oxide increases.[6,23] The Si2p$_{3/2}$ spectrum is identified to be Si-O at ~103 eV (Figure 2c),[24] while O1s signal at 532-533 eV subsumes the signals of B$_2$O$_3$ and SiO$_2$ (Figure 2d).[23,25,26] Both Si2p$_{3/2}$ and O1s spectra are not significantly modified before and after catalysis. In the XPS spectrum of the as-prepared catalyst, the N1s signals at 398.8 and 400.2 eV (Figure 2e) were assigned to B-N and N–H respectively.[23] In contrast, after catalysis, the signal at 398.8 eV is strongly reduced. This observation further confirms that BN was converted to B$_2$O$_3$ during the reaction, and since the catalyst remained stable for several hours, it suggests that B$_2$O$_3$ is also catalytically active for propane dehydrogenation. Reactions using only the DFNS support displayed poor performance, whilst pure *h*-BN showed lower conversion and productivity compared to DFNS/BN (Figure S2).

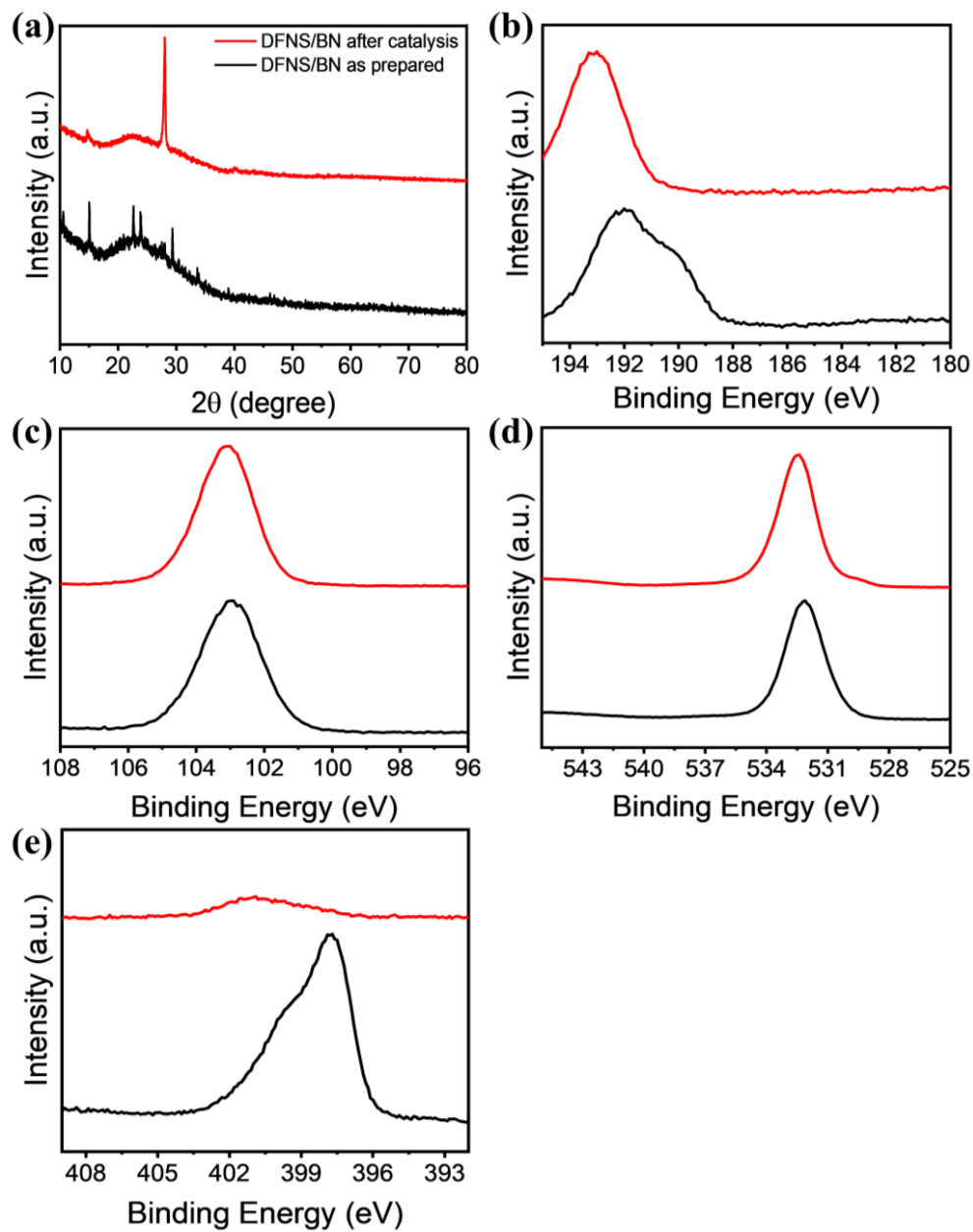

**Figure 2.** (a) PXRD and (b-e) core-level XPS spectra of (b) B1s, (c) Si2p$_{3/2}$, (d) O1s, (e) N1s elements for DFNS/BN before (black) and after (red) catalysis.

In order to test the influence of B$_2$O$_3$ on the catalytic activity for propane dehydrogenation, we prepared B$_2$O$_3$ supported on DFNS using using solid-phase synthesis deposition techniques with boric acid as a precursor. Figures 3a and b show the SEM and TEM images of DFNS/B$_2$O$_3$. As expected, they reveal that the fibrous morphology of DFNS remained intact and uniformly coated by a B$_2$O$_3$ layer, with no particle formation observed outside silica nanospheres. The B$_2$O$_3$ loading on DFNS was estimated by SEM-EDX analysis (at 10 different points to minimise the error), and was found to be 10 wt.%. Its nitrogen sorption isotherm (Figure 3c) can be classified as type IV with a type H4 hysteresis loop, associated with the presence of mesopores that are not well-defined.[18-22] The BET surface area (SA) and BJH pore volume (PV) were determined to be higher (371 $m^2.g^{-1}$ and 0.41 $cm^3.g^{-1}$ respectively), than those of DFNS/BN, possibly owing to the smaller thickness of boron oxide film with respect to BN. DFNS/$B_2O_3$ showed pore size distribution (inset Figure 3c) similar to DFNS due to its fibrous morphology, which was kept intact even after $B_2O_3$ coating. DFNS/$B_2O_3$ was then evaluated for the catalytic propane dehydrogenation reaction under exactly the same conditions as those of DFNS/BN (Figure 3d-f).

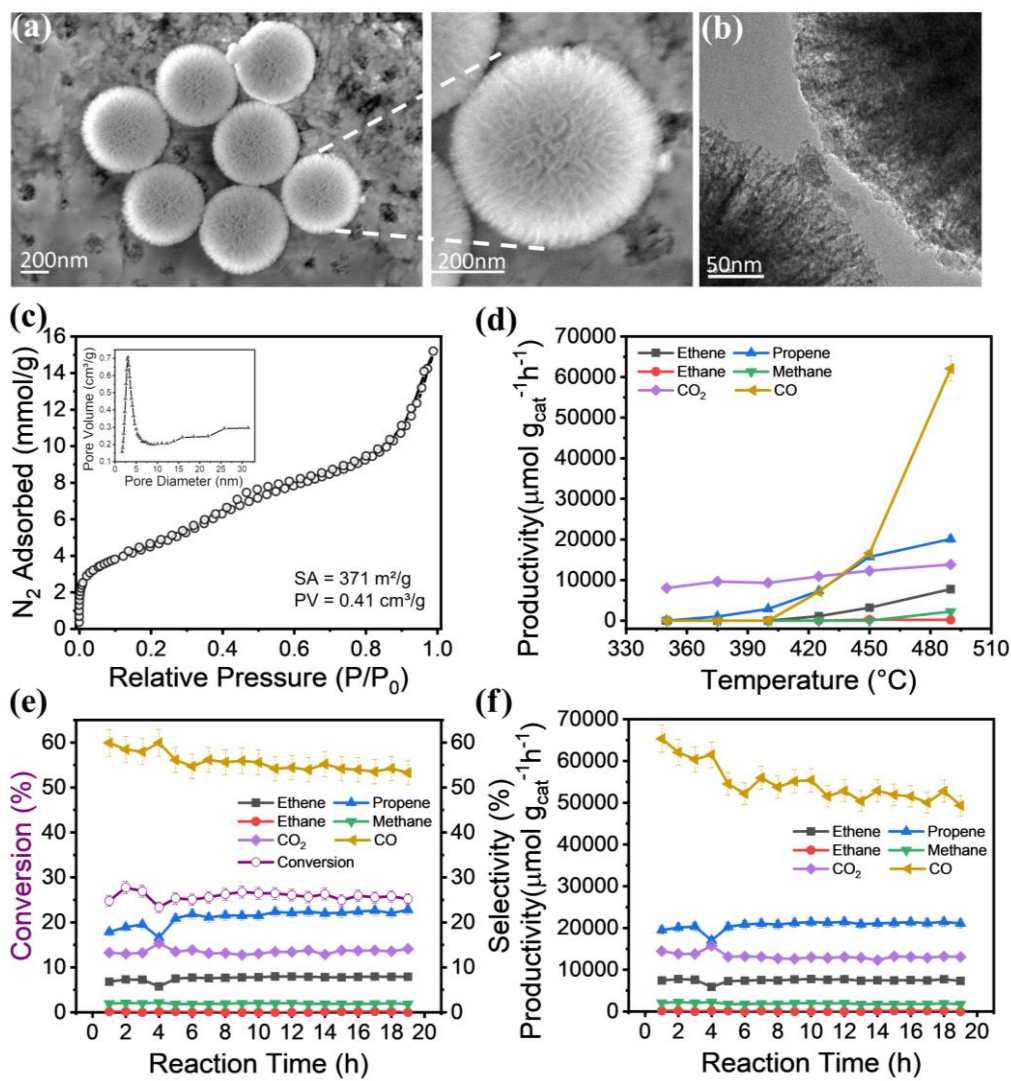

**Figure 3.** (a) SEM, (b) TEM, (c) $N_2$ sorption isotherm (pore size distribution using absorption data is shown in inset) of DFNS/$B_2O_3$. (d) Effect of temperature on productivity for propane dehydrogenation using DFNS/$B_2O_3$ catalysts. (e, f) Time evolution of (e) conversion and selectivity as well as (f) productivity for the same reaction and catalyst as in subfigure (d) carried out at 490 °C. Background in SEM images is a rough surface of the aluminum grid. Error bars were calculated by conducting reactions three times.

The catalytic activity and stability of DFNS/$B_2O_3$ for propane dehydrogenation was studied at 490 °C, in terms of conversion and selectivity (Figure 3).It is observed that the catalyst was stable for 19 h, with around 17.9% selectivity for propene and 6.8% for ethene and more than 70% for $CO_x$ with $x$ =

1 or 2, with a conversion rate of 24.8%. High selectivity towards the CO need further investigation. In terms of productivity, it showed good propene (19520 $\mu mol.g_{cat}^{-1}.h^{-1}$) and ethene (7441 $\mu mol.g_{cat}^{-1}.h^{-1}$) productivities (Figure 3f), nearly half of the DFNS/BN (Figure 1f). This indicates that $B_2O_3$ is catalytically active, albeit, to a lesser extent than BN.

PXRD and XPS analyses of DFNS/$B_2O_3$ before and after catalysis were also performed (Figure 4). The PXRD results indicate that DFNS/$B_2O_3$ (before and after catalysis) contains crystalline $B_2O_3$ phase (JCPDS-00-013-0570) (Figure 4a). XPS analysis also confirms the presence of $B_2O_3$ phase. The B1s signal at 192-193 eV in Figure 4b, was attributed to boron oxide species.[23,27] The Si2p$_{3/2}$ peak at ~103 eV in Figure 2c was assigned to Si-O,[24] while the O1s signal at 532-533 eV in Figure 2d, is the sum of contribution of $B_2O_3$ and $SiO_2$ phase.[23,25,26]

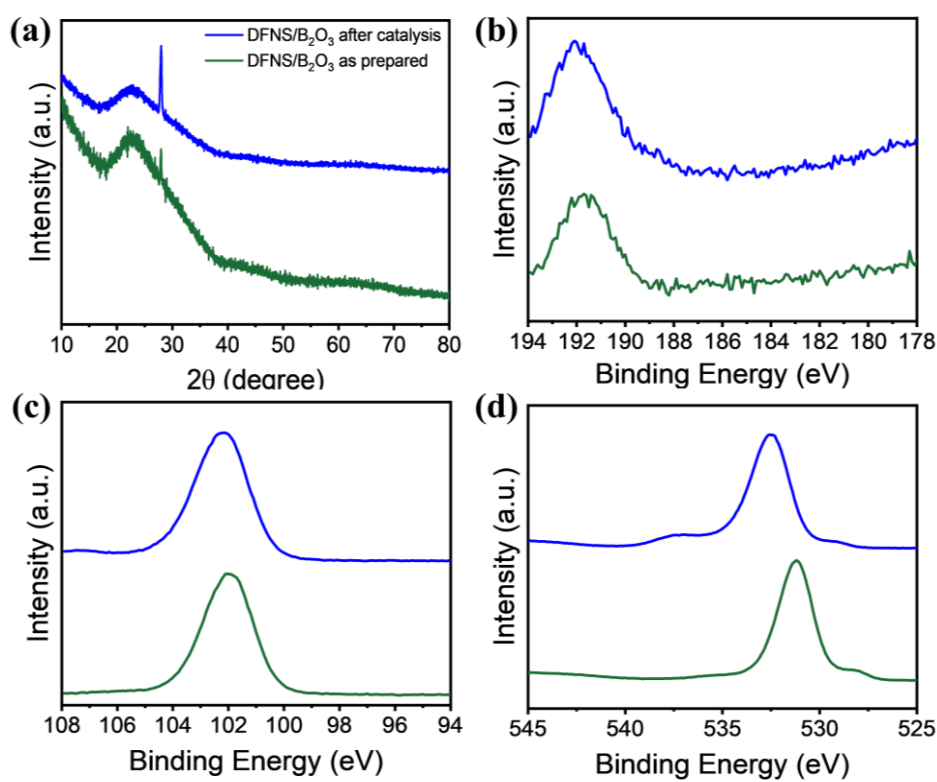

**Figure 4.** (a) PXRD and (b-d) core-level XPS spectra of (b) B1s, (c) Si2p$_{3/2}$, (d) O1s elements for DFNS/$B_2O_3$ before (green) and after (blue) catalysis.

Notably, when we compared these catalysts, DFNS/BN was found to be better than *h*-BN in terms of propane conversion and propene productivity, while BNNT showed better propene productivity compared to DFNS/BN (Figure 5). DFNS/$B_2O_3$ also displayed better performance than *h*-BN as well as a better conversion but a lower productivity than $B_2O_3$ supported on SBA-15 (BOS-10) and BNNT. Better performance of both DFNS/BN and DFNS/$B_2O_3$ was because of highly accessible active sites due to fibrous morphology and high surface area of DFNS. These DFNS/BN and DFNS/$B_2O_3$ supported catalysts showed better performance than vanadium (V) supported on silica, one of the best-known silica-supported catalysts.[5]

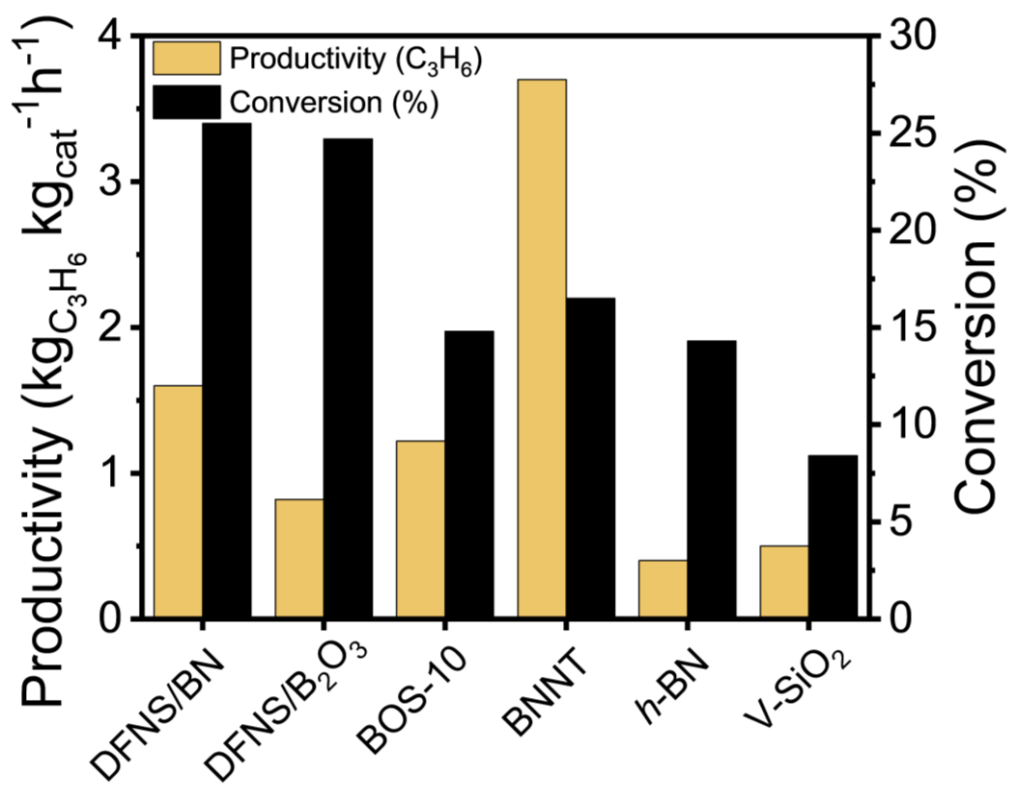

**Figure 5.** Comparison of DFNS/BN and DFNS/B$_2$O$_3$ with a selection of previously reported unsupported and supported catalysts for propane dehydrogenation.[5,12]

**Table 1.** Comparison of conventional catalysts with DFNS/BN and DFNS/B$_2$O$_3$ for propane dehydrogenation.[5,12]

| Name of Catalyst | Temperature (°C) | WHSV$^{-1}$ (kg-catsmol C$_3$H$_8^{-1}$) | Conversion % | Productivity (g $_{olefin}$ g$_{cat}^{-1}$h$^{-1}$) | Ref. |
|---|---|---|---|---|---|
| DFNS/BN | 490 | 1.47 | 25.5 | 1.9 | In this work |
| DFNS/B$_2$O$_3$ | 490 | 1.47 | 24.7 | 1.0 | In this work |
| BOS-10 | 450 | 0.18 | 14.8 | 1.2 | 78 |
| BNNT | 490 | 2-4 | 16.5 | ~4.1 | 5 |
| h-BN | 490 | 15-40 | 14.3 | ~0.4 | 5 |
| V-SiO$_2$ | 490 | 5-15 | 8.4 | ~0.5 | 5 |

The local atomic environments of the DFNS supported BN and B$_2$O$_3$ catalysts were then studied by solid-state NMR spectroscopy, a technique that has been extensively demonstrated to be a powerful probe of the local structure of materials.[28-32] Indeed, we have recently shown the successful use of solid-state NMR to better understand catalytic sites and catalytic mechanisms.[33-35] We first conducted magic-angle spinning (MAS) solid-state 1D $^1$H, $^{11}$B, and $^{29}$Si NMR experiments on DFNS/BN (as-prepared and after catalysis) and as-prepared DFNS/B$_2$O$_3$. The $^1$H NMR spectrum of as-prepared DFNS/BN is dominated by a peak at 6.8 ppm and also exhibits a broad resonance with a maximal intensity at 4.3 ppm extending down to 0 ppm. The peak at 6.8 ppm is assigned to hydroxyl groups on the edge of BN sheets.[36-38] This signal can also subsume contributions from the NB$_2$H sites of the armchair edges of BN sheets as well as BO$_3$OH moieties since the reported isotropic chemical shifts, $\delta_{iso}$, of these sites are *ca*. 6 and 8 ppm, respectively.[13,38] The broad signal centred at 4.3 ppm subsumes the contributions of adsorbed water,[39-42] BO$_2$OH sites having $\delta_{iso}$ values ranging from 4.3 to 2.7 ppm depending on the formation of hydrogen bonds and the nature of the second neighbour, B or Si,[12-14,43,44,] the NB$_2$H sites of the zigzag edges of BN sheets resonating at *ca*. 3 ppm[38] and silanol groups with $\delta_{iso}$ values ranging from 3 to 1.6 ppm depending on the formation of hydrogen bonds and the nature of the second neighbour, B or Si.[12-14,39,40,43,44] The $^1$H NMR spectrum of DFNS/BN after catalysis shown in Figure 6d displays a broad peak centred at 4.4 ppm, which subsumes the contribution of adsorbed water, BO$_2$OH sites and silanol groups. The peak at 0.1 ppm arises from coke formation or the surface residual organic species. The marked reduction in the intensity of the resonance at 6.5 ppm indicates the conversion of BN into B$_2$O$_3$ in agreement with XPS and PXRD data. The $^1$H spectrum of as-prepared DFNS/B$_2$O$_3$ shown in Fig. 6g exhibits peaks at 2.9 and 1.7 ppm, which are ascribed to BO$_2$OH and isolated silanol sites, respectively. Silanol groups close to B atoms with $\delta_{iso}$ values ranging from 2.2 to 2.7 ppm can contribute to the $^1$H signal.[14,44-46] The peak at 0 ppm is due to mobile residual organic species. The higher chemical shift of BO$_2$OH signal for DFNS/BN after catalysis than for as-prepared DFNS/B$_2$O$_3$ could result from the larger amount of hydrogen-bonded BO$_2$OH sites.[13]

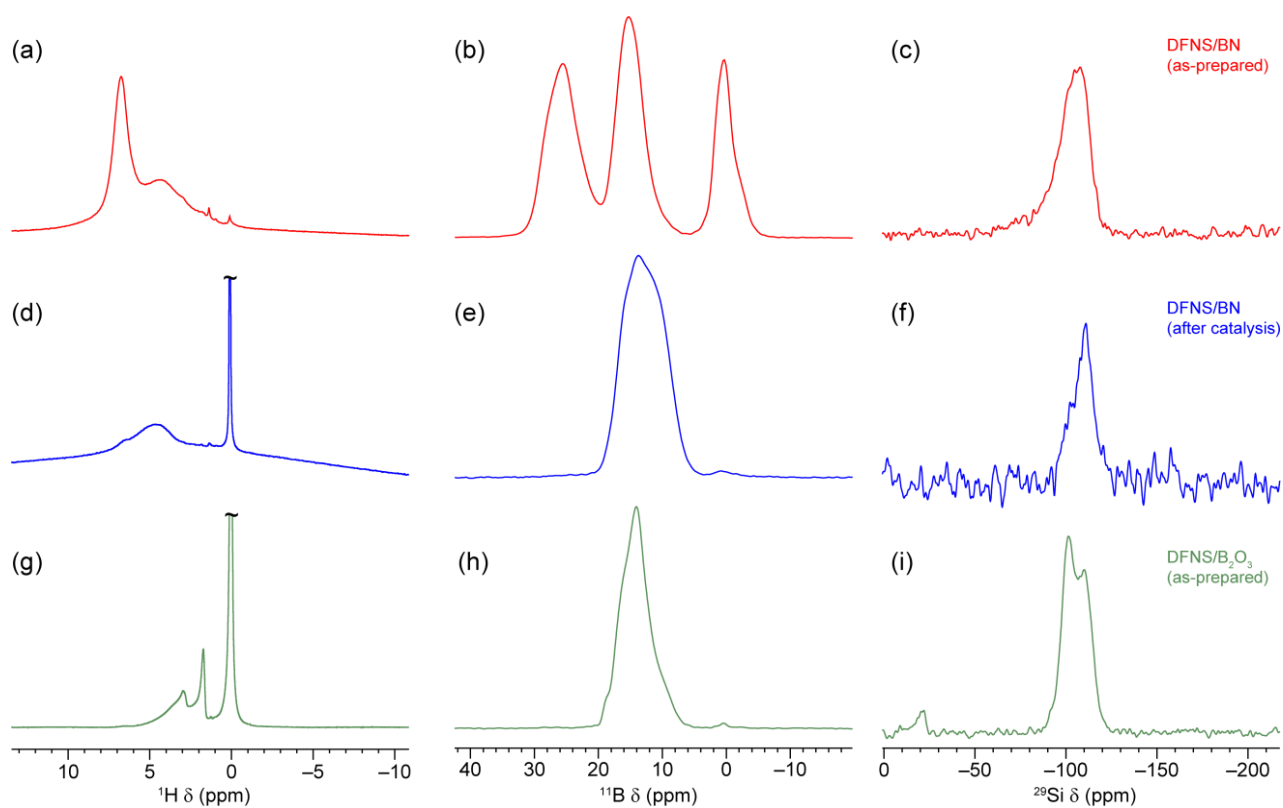
9

**Figure 6.** 1D solid-state MAS NMR spectra of (a-c) as-prepared DFNS/BN, (d-f) DFNS/BN after catalysis and (g-i) as-prepared DFNS/$B_2O_3$. (a, d, g) $^1H$ DEPTH, (b, e, h) $^{11}B$ quantitative (short tip angle), (c, f, i) $^1H \rightarrow ^{29}Si$ CP MAS. Experiments were recorded at (a, b, d, e, g, h) $B_0$ = 18.8 T with $\nu_R$ = 20 kHz and (c, f, i) $B_0$ = 9.4 T with $\nu_R$ = 10 kHz.

1D $^{11}B$ NMR data for the three aforementioned samples is presented in Figure 6b, e and h. The spectrum of as-prepared DFNS/BN (see Figure 6b) shows three broad signals with centres of gravity at *ca*. 0, 15 and 26 ppm. The signal at ~26 ppm is assigned to tri-coordinated ($B^{III}$) boron species of type $BN_3$, $BN_2O$ and $BNO_2$ since these sites have $\delta_{iso}$ values of 30, 27-28 and 20-24 ppm, respectively and quadrupolar coupling constant $C_Q$ = 2.7-2.9 MHz.[8,13,47,48] The $BN_2O$ and $BNO_2$ sites are located on the edges of the BN sheets. The signal at ~15 ppm is assigned to tri-coordinated $BO_3$ environments.[12,13,44,46,49-53] The signal at ~0 ppm is assigned to tetra-coordinated ($B^{IV}$) $BO_4$ environments.[44,46,49,50,54] The spectra of DFNS/BN after catalysis and as-prepared DFNS/$B_2O_3$ appear to be relatively similar. Both display lineshape components with centres of gravity at ~14 ppm attributed to $BO_3$ sites. This line is slightly broader for the DFNS/BN sample after catalysis. The signal at 26 ppm of $BN_3$ and $BN_2O$ sites has a very low intensity for the DFNS/BN sample after catalysis, which further confirms the oxidation of BN. As expected, this signal is not observed in the $^{11}B$ spectrum of as-prepared DFNS/$B_2O_3$. Spectra of DFNS/BN after catalysis and as-prepared DFNS/$B_2O_3$ display a small peak at ~0 ppm that can be assigned to $BO_4$ environments.

To confirm the above assignment and probe $^1H$-$^{11}B$ proximities, we recorded 2D $^{11}B$-{$^1H$} through-space heteronuclear multiple-quantum coherence (*D*-HMQC) spectra of as-prepared DFNS/BN (Figure 7a) and as-prepared DFNS/$B_2O_3$ (Figure 7b). The *D*-HMQC spectrum of as-prepared DFNS/BN is dominated by a cross-peak between hydroxyl groups on the edge of BN sheets at 6.7 ppm and $^{11}BO_3$ signal at 15 ppm. This observation is consistent with the formation of a disordered boron oxide phase at the surface of *h*-BN.[8,11] The $BO_3$ signal also correlates with the $^1H$ signal at 4 ppm assigned to $BO_2OH$ sites, which substantiates the presence of these sites in this sample. The lower intensity of the cross-peak at 4 ppm with respect to that at 6.7 ppm is consistent with the relative intensities of these signals in the 1D $^1H$ spectrum. The $^{11}B$ signals at ~0 ppm of $BO_4$ sites also exhibit weaker correlations with $^1H$ signals at 4 and 6.7 ppm, which indicates the close proximity between $BO_4$ environments and hydroxyl groups on the edge of BN sheets as well as $BO_2OH$ sites. Nevertheless, the smaller intensities of cross-peaks with $^{11}B$ signal at ~0 ppm show that hydroxyl groups are closer to $BO_3$ sites. The $^{11}B$ signal at 26 ppm exhibits further reduced correlations with $^1H$ signals at 6.7 ppm, which could stem from $BN_2(OH)$ sites at the edge of BN sheets, as well as at 4 ppm, which indicates the close proximity between $BO_2OH$ sites and BN sheets. These correlations with $^{11}B$ signal at 26 ppm are weak because they only arise from B atoms located at the edge of BN sheets. The *D*-HMQC spectrum of as-prepared DFNS/$B_2O_3$ (Figure 7b) is dominated by a correlation between BO$_3$ units and $^1$H signal ranging from 2.3 to 4 ppm, which is assigned to BO$_2$OH sites with second neighbours, B or Si, as well as silanol groups near BO$_3$ moieties.[12-14,43-45] A cross-peak between the $^1$H signal at 1.7 ppm of isolated silanol and BO$_3$ signal is also visible. Hence, the $^{11}$B-{$^1$H} *D*-HMQC spectrum confirms the close proximity between the silica surface and the boron atoms.

1D $^1$H→$^{29}$Si cross-polarisation under MAS (CP MAS) experiments were also performed on DFNS/BN, DFNS/BN after catalysis and as-prepared DFNS/B$_2$O$_3$ (see Figures 6c, f and i). Spectral fitting was carried out using DMFit[55] (with a Gaussian/Lorentzian model) in order to identify the individual species present. Detailed information on these fits can be found in the supporting information (Figure S11 and Table S2). The spectra of all three samples show the presence of Si(SiO)$_n$(OH)$_{4-n}$ sites (denoted Q$^n$) with *n* = 2, 3 and 4, characteristic of silica-type materials.[35,56,57] The spectrum of as-prepared DFNS/BN also displays a broad shoulder in the region attributable to SiN$_n$O$_{4-n}$ sites with *n* = 1, 2 or 3 (Figures 6a and S11a), since the SiN$_3$O, SiN$_2$O$_2$ and SiNO$_3$ sites resonate at *ca.* −63, −75 and −90 ppm, respectively.[58-63] This signal indicates the presence of Si-N bonds in this material. The spectrum of DFNS/BN after catalysis reveals that primarily Q$^4$ species remain in this sample, indicating that most of the Si-OH surface groups have been consumed or transformed, likely as a result of the catalytic reaction process.[64] This reduced amount of surface hydroxyls also leads to a lower efficiency of the CP transfer and hence, lower signal-to-noise ratio for this sample, despite the collection of a larger number of transients.[65,66] The shoulder centred at *ca.* −73 ppm (attributed to SiN$_n$O$_{4-n}$ sites) is also no longer visible after catalysis, in agreement with the conversion of BN into B$_2$O$_3$ inferred from PXRD, XPS and $^{11}$B NMR.

To further probe the interactions between the silica surface and the supported BN and B$_2$O$_3$ phases, we record 2D $^{11}$B-{$^{29}$Si} *D*-HMQC spectra. The corresponding spectrum of as-prepared DFNS/BN shown in Figure 7c is dominated by a cross-peak between the Q$^4$ site and the BO$_3$ signal at ~12 ppm assigned to B(OSi)$_3$ environments.[12,14,45,49] These B(OSi(Q$^4$))$_3$ sites mainly corresponds to the subsurface region of the material. The spectrum also exhibit cross-peaks between the Q$^4$ and Q$^3$ site and the BO$_3$ signal at ~14 ppm as well as between the Q$^3$ sites and the BO$_3$ signal at ~16 ppm, which are assigned to B(OSi(Q$^{3,4}$))$_2$OH and BOSi(Q$^3$)(OH)$_2$ surface sites, respectively.[14,45,49] We also observe cross-peaks between the BO$_4$ signal at ~0 ppm and the Q$^3$ sites as well as the BO$_4$ signal at ~−2 ppm and the Q$^4$ sites. The cross-peak of the B$^{IV}$ signal at ~0 ppm can be ascribed to the surface B(OSi(Q$^3$))$_3$OH site, whereas the cross-peak of the B$^{IV}$ signal at ~−2 ppm can be ascribed to the subsurface B(OSi(Q$^4$))$_4$ sites.[13,54] These cross-peaks between BO$_3$ and BO$_4$ signals and those of Q$^4$ and Q$^3$ sites demonstrate the presence of B−O−Si bonds in the as-prepared DFNS/BN. Furthermore, the BN$_2$OH signal at 26 ppm correlates with that of SiN$_2$O$_2$ site at −73 ppm. We can also observe a weak cross-peak between the BNO$_2$ sites at 23.5 ppm and the SiNO$_3$ signal at −93 ppm.[38,59,61,63] These cross-peaks suggest the presence of B−N−Si bonds in the sample. The 2D $^{11}$B-{$^{29}$Si} *D*-HMQC spectrum of DFNS/$B_2O_3$ (Figure 7d) is dominated by cross-peaks between the $BO_2OH$ signal at ~14 ppm and the $Q^4$ and $Q^3$ sites, which suggests the anchoring of $B_2O_3$ phase on silica surface *via* B−O−Si bonds.[12]

To better resolve the $^{11}B$ signals, we also acquired 2D $^{11}B$ multiple-quantum magic-angle spinning (MQMAS) spectra of as-prepared DFNS/BN and DFNS/$B_2O_3$ (Figure 7e and f) and DFNS/BN after catalysis (see Figure S7c and e). These spectra correlate the anisotropic MAS spectra of the $^{11}B$ central transitions shown along the direct $\delta_2$ dimension to their isotropic shift detected along the indirect $\delta_1$ dimension. The projection along the vertical isotropic direction indicates that each of the three $^{11}B$ sites has a wide distribution of $\delta_{iso}$ values, which stems from the distribution of local environments, as the second neighbour of boron atoms can be B, Si or H.[44,49]

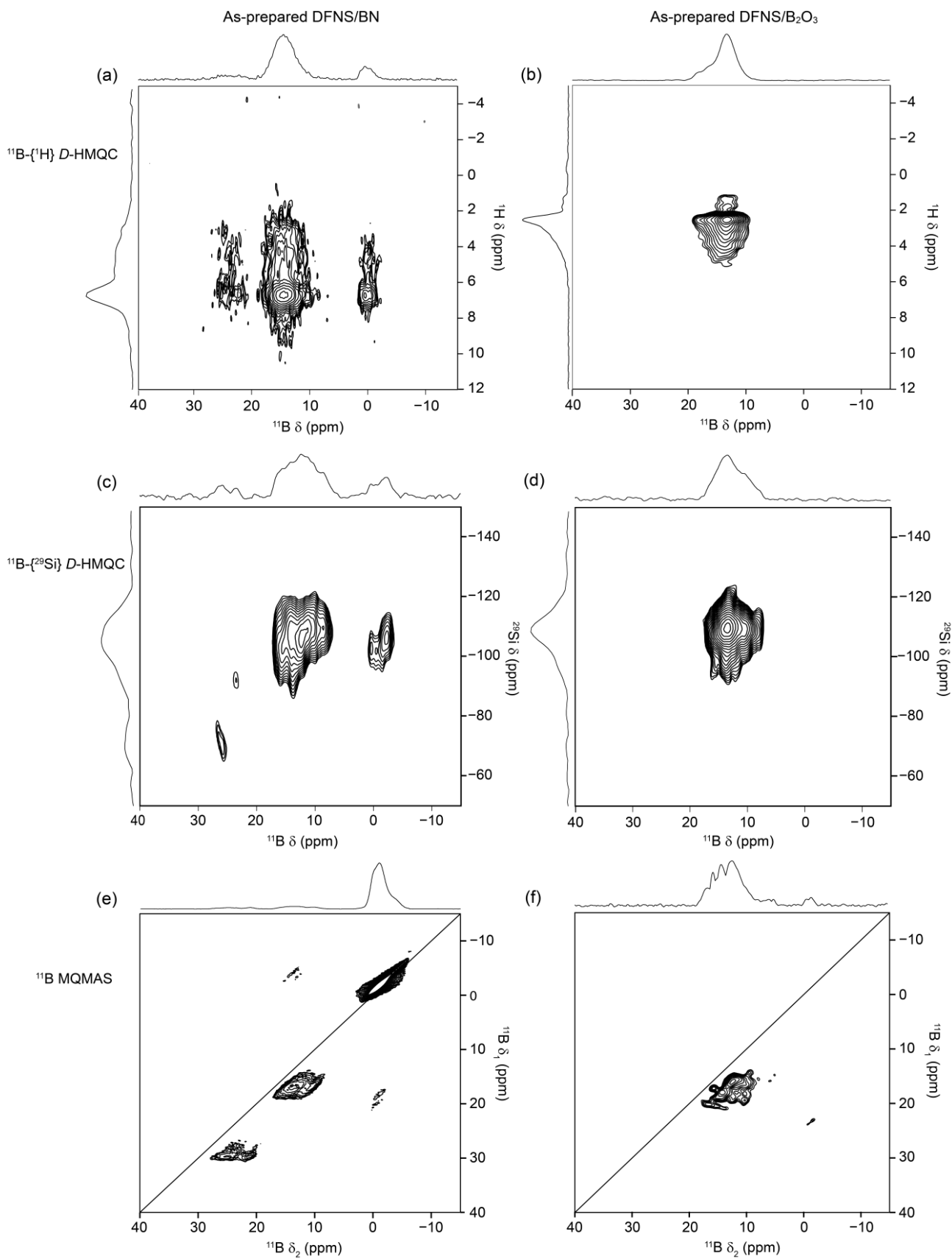

**Figure 7.** (a, b) 2D $^{11}B$-$\{^{1}H\}$ *D*-HMQC, (c, d) 2D $^{11}B$-$\{^{29}Si\}$ *D*-HMQC NMR and (e, f) 2D $^{11}B$ MQMAS spectra of (a, c, e) as-prepared DFNS/BN and (b, d, f) as-prepared DFNS/$B_2O_3$. The MQMAS and $^{11}B$-$\{^{1}H\}$ *D*-HMQC spectra were acquired at $B_0$ = 18.8 T and $\nu_R$ = 20 kHz, whereas the $^{11}B$-$\{^{29}Si\}$ *D*-HMQC NMR spectra were recorded at $B_0$ = 18.8 T and $\nu_R$ = 14.286 kHz.

In order to obtain detailed insight into the distinct functional groups that were present in the boron surface layers, spectral fitting of the lineshapes in the $F_1$ dimension of the acquired MQMAS spectra was performed using DMFit.[55] Since the $F_1$ dimension of an MQMAS spectrum presents resonances that are free from the effects of quadrupolar broadening,[31,55] it was possible for a Gaussian/Lorentzian model to be employed. The simulated spectra are shown in Figures S4 to S6. Table S1 lists the shifts, $\delta_1$, along the indirect dimension of the sheared MQMAS spectra of the simulated peaks for the three investigated samples. Due to the significant overlap of sites in these MQMAS spectra, fitting of the total lineshapes to obtain $\delta_{iso}$ and $C_Q/P_Q$ values with a sufficient degree of accuracy was not possible.

In the 2D MQMAS spectrum of as-prepared DFNS/BN, three peaks at $\delta_1$ = 30.5, 28.6 and 27.2 ppm can be resolved in the most deshielded signals. The peak at 30.5 ppm is assigned to bulk $BN_3$ sites, whereas the signals at 28.6 and 27.2 ppm are likely produced by $BN_2OH$ and $BN_2O$ sites, respectively, located at the edge of BN sheets, since the $^{11}B$ chemical shift increases for decreasing number of bridging oxygen atoms.[38,53,67] The $BNO_2$ signal may not be detected because of the lack of sensitivity of MQMAS experiments. The $BO_3$ signal can be simulated by three peaks at $\delta_1$ = 18.4, 17.2 and 16.3 ppm, assigned to $BO_2OH(ring)$, $BO_2OH(non-ring)$ and $BO_3(ring)$, respectively since for the same number of bridging oxygens, $^{11}B$ nuclei residing in boroxol rings exhibit more deshielded signals as compared to those in the chain.[12,53,67] $BO_3(non-ring)$ sites resonating at $\delta_1 \approx 14$ ppm (see below) can also contribute to the signal at $\delta_1$ =16.3 ppm. Furthermore, according to the 2D $^{11}B$-{$^{29}Si$} D-HMQC spectrum, a fraction of the $BO_3$ moieties are anchored to the silica surface. In particular, the peak at $\delta_1$ = 18.4 ppm subsumes the signal of $BOSi(Q^3)(OH)_2$ and $B(OSi(Q^{3,4}))_2OH$ surface sites, whereas the signal of $B(OSi(Q^4))_3$ sites also overlaps with those of $BO_2OH(non-ring)$ and $BO_3(ring)$ environments. The $BO_4$ signal can be deconvoluted as the sum of three peaks at $\delta_1$ = 0.1, −1 and −1.9 ppm, which may be ascribed to $B(OSi(Q^3))_3OH$, $B(OB)_4$ and $B(OSi(Q^4))$ sites, respectively (see Fig. S2), based on the 2D $^{11}B$-{$^1H$} and $^{11}B$-{$^{29}Si$} D-HMQC spectra (see Fig. 7a,c).

As seen in Figure S5, the $BO_3$ signal in the projection of the 2D MQMAS spectrum of DFNS/$B_2O_3$ can be simulated as the sum of five components resonating at $\delta_1$ = 20.5, 19, 17.2, 15.8 and 14.4 ppm, assigned to $BO(OH)_2$, $BO_2OH(ring)$, $BO_2OH(non-ring)$, $BO_3(ring)$ and $BO_3(non-ring)$, respectively.[12,53,67] The projection also displays a folded $BO_4$ signal, which can be simulated as a single resonance. For DFNS/BN after catalysis, the $BO_3$ signal in the projection of the 2D MQMAS spectrum can be simulated as the sum of five components at $\delta_1$ = 21.1, 19.1, 17.1, 15.7 and 14.1 ppm (see Fig. S6). These shifts are similar to those used to simulate the $BO_3$ signal DFNS/$B_2O_3$. The $BO_4$ signal of DFNS/BN after catalysis can also be simulated as a single resonance. A key difference between DFNS/$B_2O_3$ and DFNS/BN after catalysis is the larger relative intensities of $BO(OH)_2$ and $BO_2OH(ring)$ sites as compared to $BO_3(ring)$ for the latter sample, which is consistent with the larger amount of hydrogen bonded hydroxyl groups attached to B for this material (see above).[13,14] This structural difference could explain the difference in productivity between DFNS/BN after catalysis and DFNS/$B_2O_3$.

## 3. Conclusions

In conclusion, we observed that DFNS/BN displayed higher catalytic activity compared to unsupported *h*-BN, V-$SiO_2$ and BOS-10 in terms of conversion and productivity of olefin. DFNS/$B_2O_3$ also showed higher activity for propane ODH but lower propene productivity. The catalytic activity of DFNS/BN depends on a complex interplay between the types and the concentrations of catalytic sites (increased reactivity due to Si-O-B bonding), fibrous morphology (provided better diffusion) and the textural properties (high surface area and pore volume).

Through a combination of analytical techniques (PXRD, XPS and multinuclear solid-state NMR spectroscopy), we have shown that the supported BN phase is oxidised to $B_2O_3$ during the propane dehydrogenation process. In our case, we did not observe any deactivation of DFNS/$B_2O_3$ up to 19 h, due to the stabilizing effect of DFNS through covalent linkage of silica and $B_2O_3$ via Si-O-B bonding, the presence of which was confirmed by solid-state NMR spectroscopy. This allowed $B_2O_3$ to remain on the DFNS surface and act as a catalyst, albeit with a more moderate selectivity than DFNS/BN. Furthermore, solid-state NMR spectra indicates that oxidised DFNS/BN exhibits from a larger amount of hydrogen bonded hydroxyl groups attached to B than as-prepared DFNS/$B_2O_3$. This structural difference could explain the higher selectivity towards propene of oxidised DFNS/BN.

**Acknowledgements**

This work was supported by the Department of Atomic Energy (DAE), Government of India (Grant no. 12P0154). V.P. and R.B. also thank the Department of Science and Technology (DST), Government of India (Grant No. CEFIPRA 5208-2) and SHELL (17R002) for funding. Institut Chevreul (FR-2638), Ministère de l'Enseignement Supérieur, de la Recherche et de l'Innovation, Région Hauts-de-France and FEDER/ERDF are acknowledged for supporting and partially funding this work. Financial support from the IR-RMN-THC FR-3050 CNRS for conducting the research is gratefully acknowledged. This project has received funding from the European Union's Horizon 2020 research and innovation program under grant agreement No. 731019 (EUSMI). OL acknowledges financial support from Institut Universitaire de France (IUF).

**Conflict of interest:** The authors declare no conflict of interest.

**Keywords:** Heterogeneous catalysis • Nanostructures • NMR spectroscopy • Propane dehydrogenation • X-ray diffraction • X-ray photoelectron spectroscopy • Electron microscopy

# Supporting Information

# Boron Nitride and Oxide Supported on Dendritic Fibrous Nanosilica for Catalytic Oxidative Dehydrogenation of Propane

Rajesh Belgamwar,[1] Andrew G. M. Rankin,[2,#] Ayan Maity,[1] Amit Kumar Mishra,[1] Jennifer S. Gómez,[2] Julien Trébosc,[2,4] Chathakudath P. Vinod,[3] Olivier Lafon,[2,5]* Vivek Polshettiwar[1]*

[1]Department of Chemical Sciences, Tata Institute of Fundamental Research (TIFR), Mumbai, India. *Email: vivekpol@tifr.res.in
[2]Univ. Lille, CNRS, Centrale Lille, Univ. Artois, UMR 8181, UCCS – Unité de Catalyse et Chimie du Solide, F-59000 Lille, France. *Email: olivier.lafon@univ-lille.fr
[3] Catalysis & Inorganic Chemistry Division, CSIR-National Chemical Laboratory (NCL), Pune, India
[4]Univ. Lille, CNRS-FR2638, Fédération Chevreul, F-59000 Lille, France
[5]Institut Universitaire de France (IUF)
[#]Present Address: Sorbonne Université, CNRS, Collège de France, Laboratoire de Chimie de la Matière Condensée de Paris (LCMCP), 4 place Jussieu, F-75005 Paris, France

**Experimental Section**

**1-1. Catalysts Synthesis**

**Synthesis of DFNS/BN.** DFNS (2 g) and water (30 mL) were added to a 250 mL conical flask and sonicated for 30 min. In another 250 mL conical flask, boric acid (1 g) and urea (24 g) were added to 30 mL of water with constant stirring and the solution was stirred at 85 °C for 15 min. To this solution, the DFNS dispersion was slowly added and then stirring was continued at 85 °C until all the solvent was evaporated. The resulting dry powder was ground using a mortar and pestle and then heated in a tubular furnace at 900 °C at a rate of 5 °C.min$^{-1}$ for 5 h. Heating was carried out under a nitrogen flow of 200 mL.min$^{-1}$ from 25 to 200 °C, and then under a N$_2$ flow of 20 mL.min$^{-1}$ from 200 to 900 °C. The resulting white powder was stored under ambient conditions.

**Synthesis of DFNS/B$_2$O$_3$.** DFNS (1 g) and boric acid (0.1 g) were mixed and ground using a mortar and pestle for 15 min. This powder was transferred to an alumina pot in a furnace, which was heated at 130 °C for 30 min and then at 330 °C for 60 min with a heating rate of 5 °C.min$^{-1}$, in air. The resulting white powder was stored under ambient conditions.

**1-2. Characterisation**

SEM images were recorded using a ZEISS ULTRA scanning electron microscope operated at an accelerating voltage of 3 - 5 kV. SEM samples were prepared by drop-casting diluted ethanolic suspension of the powder onto the aluminium stub. TEM images were recorded with a FEI-TECNAI transmission electron microscope operated at an accelerating voltage of 200 kV. TEM samples were prepared by drop-casting a diluted ethanolic suspension of the powder onto the porous carbon copper grid (300 mesh), in order to avoid agglomeration. Elemental analysis was measured by energy-dispersive X-ray spectroscopy (EDS). Brunauer-Emmet-Teller (BET) Nitrogen adsorption measurements were performed using a Micromeritics 3-Flex analyser. The tubes were filled with approximately 100 mg of powder sample and then degassed for 12 h at 120 °C in an external degasser. The tubes were then attached to the analyser, and prior to starting the analysis, were degassed again at 120 °C for 2 h to ensure the removal of all moisture and adsorbed gases. X-ray diffraction patterns were recorded using a Panalytical X'Pert Propowder X-ray diffractometer employing Cu-Kα radiation.

### 1-3. Catalytic Propane Dehydrogenation Reaction

The propane dehydrogenation reaction was carried out in a fixed bed continuous flow reactor. A volume of 200 mg of catalyst were placed in a tubular reactor with 0.9 cm inner diameter and 12 cm length. The reactor was heated to 490 °C at a rate of 10 °C.min$^{-1}$ with 12 mL.min$^{-1}$ flow of $N_2$ for 30 min. Then a mixture of reactive gases was fed (at a rate of 2 mL.min$^{-1}$ $O_2$, 2 mL.min$^{-1}$ propane and 8 mL.min$^{-1}$ $N_2$ as an internal standard) by mass flow controllers (MFC) for individual gases. Monitoring of the reaction progress and product quantification was carrried out using a gas chromatography instrument (GC, Agilent 7890 B) equipped with a flame ionisation detector (FID) and a thermal conductivity detector (TCD), using an Agilent hybrid column CP7430 to separate the $CO_2$, $CH_4$ and CO.

Using standard calibration gases, we calculated the propane, propene, ethane, ethene, methane, $CO_2$ and CO production in ppm, using the respective GC peak areas. The propane conversion was estimated as

$$\text{Propane Conversion (\%)} = \frac{(\text{Propane } \mu\text{mol.min}^{-1})_{in} - (\text{Propane } \mu\text{mol.min}^{-1})_{out}}{(\text{Propane } \mu\text{mol.min}^{-1})_{in}} \times 100 \quad (S1)$$

and the propene selectivity was determined as

Propene Selectivity (%) =

$$\frac{(\text{Propene})_{\mu\text{mol.min}^{-1}}}{(\text{Propene})_{\mu\text{mol.min}^{-1}} + (\text{Ethene})_{\mu\text{mol.min}^{-1}} + (\text{Ethane})_{\mu\text{mol.min}^{-1}} + (\text{Methane})_{\mu\text{mol.min}^{-1}} + (\text{CO})_{\mu\text{mol.min}^{-1}} + (\text{CO}_2)_{\mu\text{mol.min}^{-1}}} \times 100 \quad (S2)$$

The selectivity for other products was also calculated using a similar equation.

The actual gas flow was estimated as

$$\text{Actual gas flow (mL.min}^{-1}) = \frac{(N_2 \text{ area})_{out \text{ in control experiment with DFNS}} \times (\text{Gas Flow mL.min}^{-1})_{total}}{(N_2 \text{ area})_{out \text{ at } 490 \text{ °C}}} \quad (S3)$$

whereas the production rate in μmol.min$^{-1}$ was obtained from the equation below

$$\text{Production rate (μmol. min}^{-1}) = \frac{(\text{Product})_{ppm} \times (\text{Actual gas flow mL.min}^{-1})}{(22400)_{cm^3.mol^{-1}}} \qquad (S4)$$

The production rate was also normalized to the mass of catalyst as

$$\text{Production rate (μmol. g}^{-1}.\text{h}^{-1}) = \text{Production rate (μmol . min}^{-1}) \times 60 \times \frac{1}{(\text{mass of catalyst in g})} \qquad (S5)$$

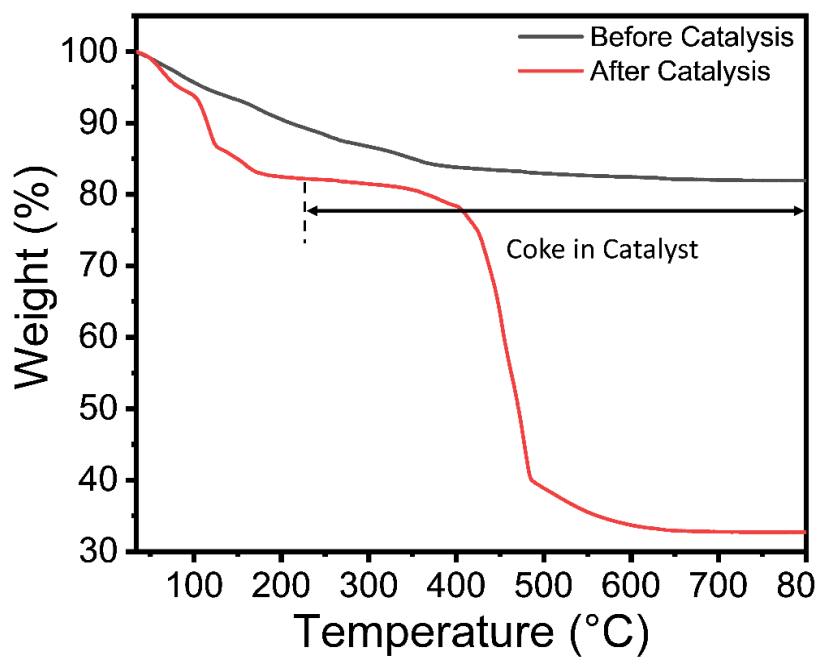

**Figure S1**. Thermal gravimetric analysis (TGA) for coke formation before and after catalysis.

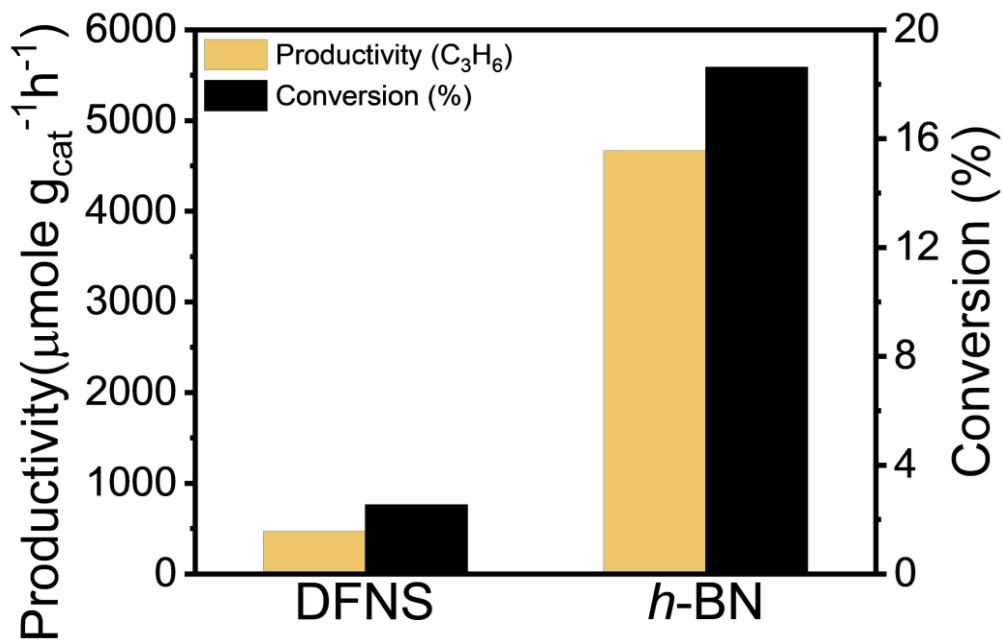

**Figure S2**. ODH reaction of propane carried out at 490 °C using DFNS and *h*-BN as catalysts.

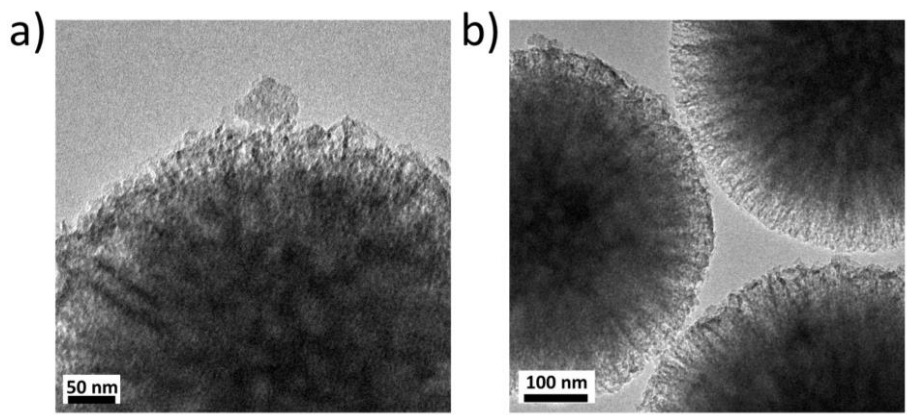

**Figure S3**. a) TEM of DFNS/BN after catalysis b) TEM of DFNS/$B_2O_3$ after catalysis.

## 1-4. Solid-state NMR spectroscopy

For all solid-state NMR experiments, rotors were packed inside an argon-atmosphere glovebox to prevent contact with moisture. $^{11}$B chemical shifts were referenced using a solid sample of NaBH$_4$ ($\delta_{iso}$ = −42.05 ppm), whereas $^1$H and $^{29}$Si chemical shifts were referenced using neat tetramethylsilane (TMS) ($\delta_{iso}$ = 0 ppm).

### 1-4.1. Experiments at 9.4 T

$^1H\rightarrow{}^{29}Si$ CP MAS NMR experiments were recorded at $B_0$ = 9.4 T ($\nu_0$ = 400 MHz for $^1H$) with a wide-bore magnet equipped with a Bruker AVANCE II console, using a 3.2 mm triple-resonance HXY magic angle spinning (MAS) probe operated in double resonance mode. The samples were spun at a MAS rate, $\nu_R$ = 10 kHz. The initial π/2 pulse lasted 2 μs and was followed by a CP contact time equal to 8 ms. During the CP transfer, the RF nutation frequency on the $^{29}Si$ channel was constant and equal to 50 kHz, while the $^1H$ nutation frequency was linearly ramped from 36 to 40 kHz. The $^1H$-$^{29}Si$ CP MAS spectra were recorded using recycle intervals of 1.95 s (as-prepared DFNS/BN), 17 s (as-prepared DFNS/B$_2$O$_3$) or 27.29 s (DFNS/BN after catalysis) and were the result of averaging 4096 (as-prepared DFNS/BN and DFNS/B$_2$O$_3$) or 7200 transients (DFNS/BN after catalysis). SPINAL $^1H$ decoupling[1] (~125 kHz) was applied during acquisition. The Hartmann-Hahn spinning sideband match condition[2,3] was calibrated using a powdered sample of octakis(trimethylsiloxy)silsesquioxane (Q8M8).

### 1-4.2. Experiments at 18.8 T

All experiments at $B_0$ = 18.8 T ($\nu_0$ = 800 MHz for $^1H$) were performed with a narrow-bore magnet equipped with a Bruker AVANCE NEO console. One-dimensional (1D) $^1H$ and $^{11}B$ experiments were acquired at $\nu_R$ = 20 kHz, using 3.2 mm double-resonance BP (featuring an MgO stator block to prevent $^{11}B$ background signal) and 3.2 mm double-resonance HX probes for $^{11}B$ and $^1H$ experiments, respectively. $^1H$ NMR spectra were acquired by averaging 16 transients separated by a recycle interval of 3 s, using the DEPTH pulse sequence for probe background suppression,[4] with $\nu_1 \approx 93$ kHz. Quantitative $^{11}B$ NMR spectra were acquired by using single-pulse NMR experiments with a pulse length of 0.5 μs and an RF field strength of 71 kHz, corresponding to a short tip angle.[5] The quantitative $^{11}B$ NMR spectra were acquired by averaging 2 (pure hexagonal (h) BN), 4096 (as-prepared DFNS/BN and DFNS/B$_2$O$_3$) or 16 (DFNS/BN after catalysis) transients, separated by a recycle interval of 1 (as-prepared DFNS/BN and DFNS/B$_2$O$_3$), 0.8 (DFNS/BN after catalysis) or 586 s (pure h-BN).

Two-dimensional (2D) $^{11}B$ MQMAS spectra were obtained using the z-filtered sequence[6] at $\nu_R$ = 20 kHz, with a 3.2 mm HX probe. Excitation and conversion pulses lasted $\tau_p$ = 6.5 and 2 μs with $\nu_1 \approx 100$ kHz and the central transition (CT) selective pulse lasted 12.5 μs with $\nu_1 \approx 20$ kHz. SPINAL $^1H$ decoupling[1] (~100 kHz) was applied during acquisition. Spectra are the result of averaging 240 (as-prepared DFNS/BN), 1032 (as-prepared DFNS/B$_2$O$_3$) or 1704 (DFNS/BN after catalysis) transients separated by a recycle delay of 1 (as-prepared DFNS/B$_2$O$_3$ and DFNS/BN after catalysis) or 3.5 s (as-prepared DFNS/BN) for each of the 256 (as-prepared

DFNS/BN) or 32 (as-prepared DFNS/B$_2$O$_3$ and DFNS/BN after catalysis) $t_1$ increments of 25 (as-prepared DFNS/BN) or 50 μs (as-prepared DFNS/B$_2$O$_3$ and DFNS/BN after catalysis). The quadrature detection method of States *et al.*[7] was used to achieve sign discrimination in the indirect dimension. After acquisition, a 2D Fourier transformation followed by a shearing transformation was performed to obtain a spectrum where the contour ridges were parallel to $F_2$, allowing the isotropic spectrum to be obtained directly from a projection onto $F_1$.[8]

$^{11}$B-{$^1$H} *D*-HMQC experiments with $^{11}$B detection were acquired at $\nu_R$ = 20 kHz with a 3.2 mm triple-resonance HX probe.[9,10] The $^1$H-$^{11}$B dipolar couplings were reintroduced by applying SR4$_1^2$ scheme on the $^1$H channel.[11] The total duration of the two SR4$_1^2$ recoupling blocks was $\tau_D$ = 0.75 ms. The $^1$H RF nutation frequencies of the π/2 pulses and SR4$_1^2$ recoupling were equal to $\nu_1 \approx$ 100 and 40 kHz, respectively. $^{11}$B CT selective pulses were applied with $\nu_1 \approx$ 20 kHz. Spectra are the result of averaging 16 (as-prepared DFNS/BN), 32 (as-prepared DFNS/B$_2$O$_3$) or 128 (DFNS/BN after catalysis) transients separated by a recycle interval of 2 (as-prepared DFNS/BN and DFNS/BN after catalysis) or 1 s (as-prepared DFNS/B$_2$O$_3$) for each of the 46 (as-prepared DFNS/BN), 100 (as-prepared DFNS/B$_2$O$_3$) or 44 (DFNS/BN after catalysis) $t_1$ increments of 50 μs. The States-TPPI quadrature detection method[12] was used to achieve sign discrimination in the indirect dimension.

2D $^{11}$B-{$^{29}$Si} *D*-HMQC experiments were acquired with a 3.2 mm triple-resonance HXY triple-gamma probe at $\nu_R$ = 14.286 kHz. The $^{11}$B-$^{29}$Si dipolar couplings were reintroduced the SFAM$_1$ dipolar recoupling scheme on the $^{29}$Si channel.[13] This experiment was calibrated using a sample of $^{29}$Si enriched borosilicate glass.[14] The total duration of the two SFAM$_1$ recoupling blocks was $\tau_D$ = 2 ms. $^{29}$Si π/2 pulses were applied with $\nu_1 \approx$ 59 kHz and $^{11}$B CT selective pulses were applied with $\nu_1 \approx$ 17 kHz. The peak values of the frequency and amplitude sweeps of SFAM$_1$ on $^{29}$Si channel were $(\nu_{ref}^{max}, \nu_1^{max})$ = (70, 57) kHz. The double frequency sweep (DFS) method[15] was used in order to enhance the polarisation of the central transition (CT) by manipulation of the populations of the satellite transitions (ST). The DFS pulse lasted 2 ms and used $\nu_1 \approx$ 17 kHz. During the DFS pulse, the frequencies of the rf spikelets were linearly swept in a symmetric manner from 1 to 85.8 kHz with respect to the CT. Spectra are the result of averaging 448 (DFNS/BN) or 1024 (DFNS/B$_2$O$_3$) transients separated by a recycle interval of 4 s for each of the 32 (DFNS/BN) or 42 (DFNS/B$_2$O$_3$) $t_1$ increments of 35 μs. The States-TPPI quadrature detection method[12] was used to achieve sign discrimination in the indirect dimension. Data processing was carried out using Bruker TopSpin (version 4.0.8) and ssNake (version 1.1).[16]

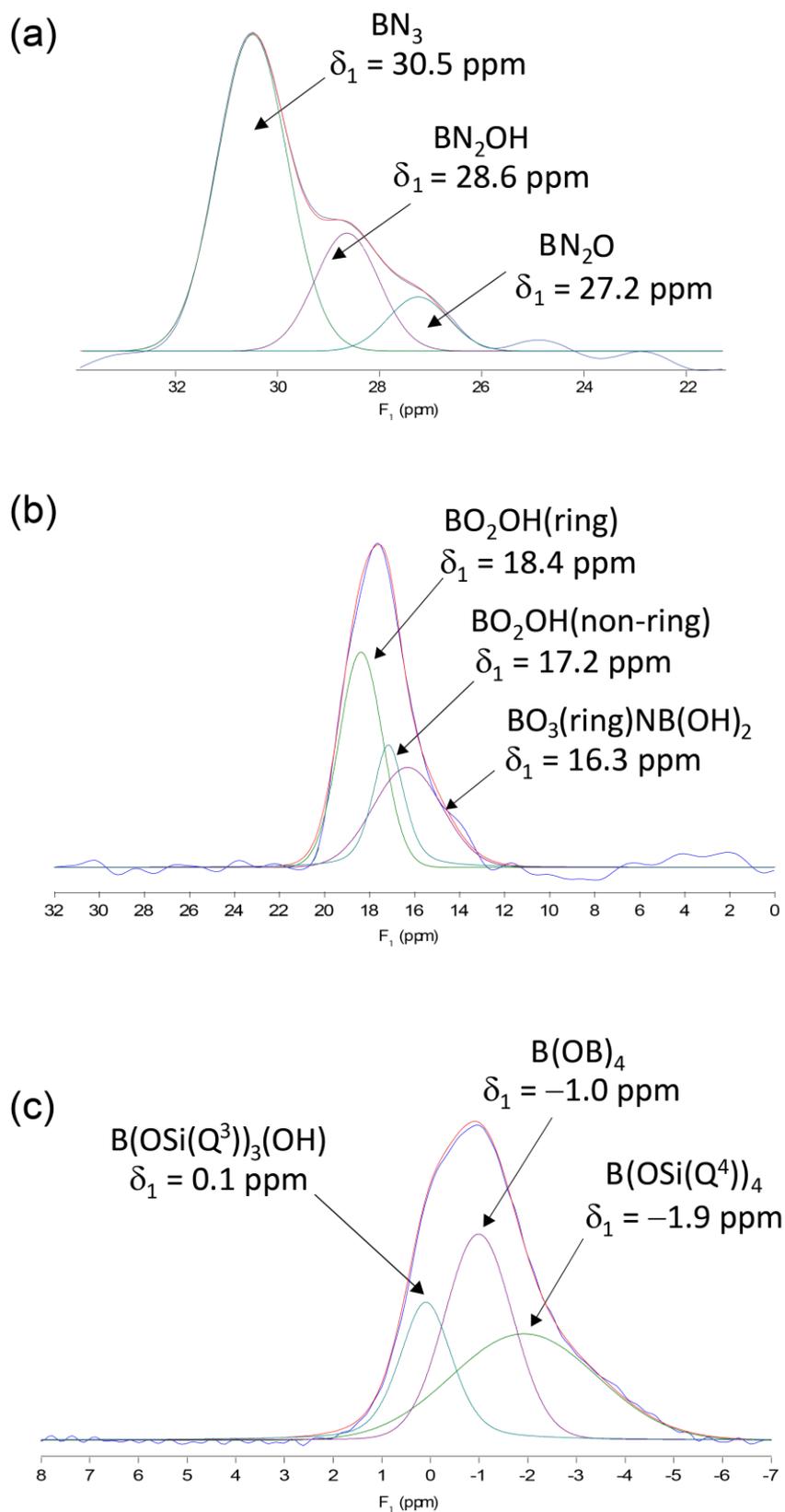

**Figure S4**. (a-c) Deconvolutions of the lineshapes obtained from total projections of the $F_1$ dimension of the $^{11}B$ MQMAS spectrum of as-prepared DFNS/BN. Fitting was carried out using the Gaussian/Lorentzian model in DMFit.[17]

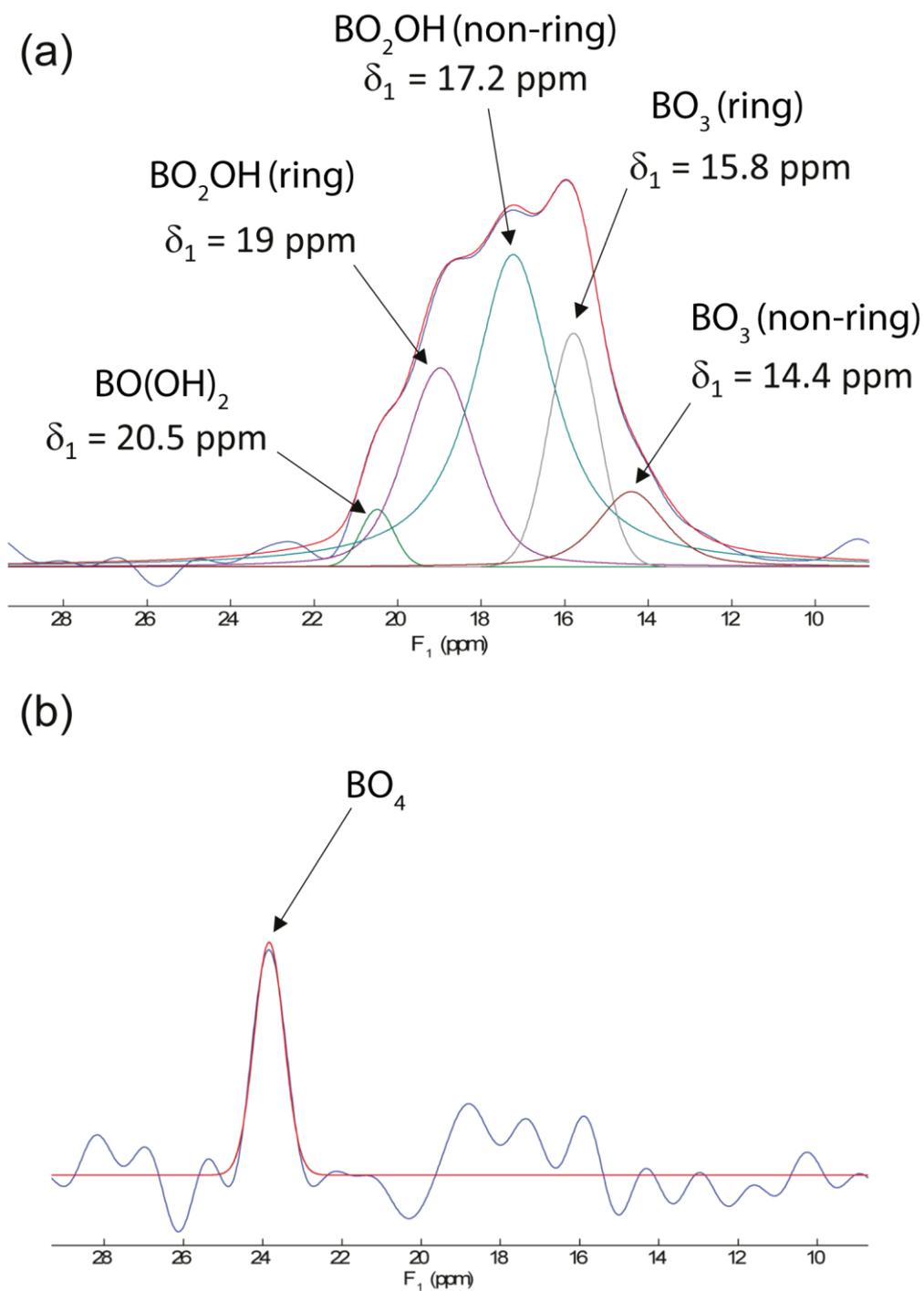

**Figure S5**. (a-b) Deconvolutions of the lineshapes obtained from total projections of the $F_1$ dimension of the $^{11}B$ MQMAS spectrum of as-prepared DFNS/$B_2O_3$. Fitting was carried out using the Gaussian/Lorentzian model in DMFit.[17] The $BO_4$ peak in the subfigure b is folded.

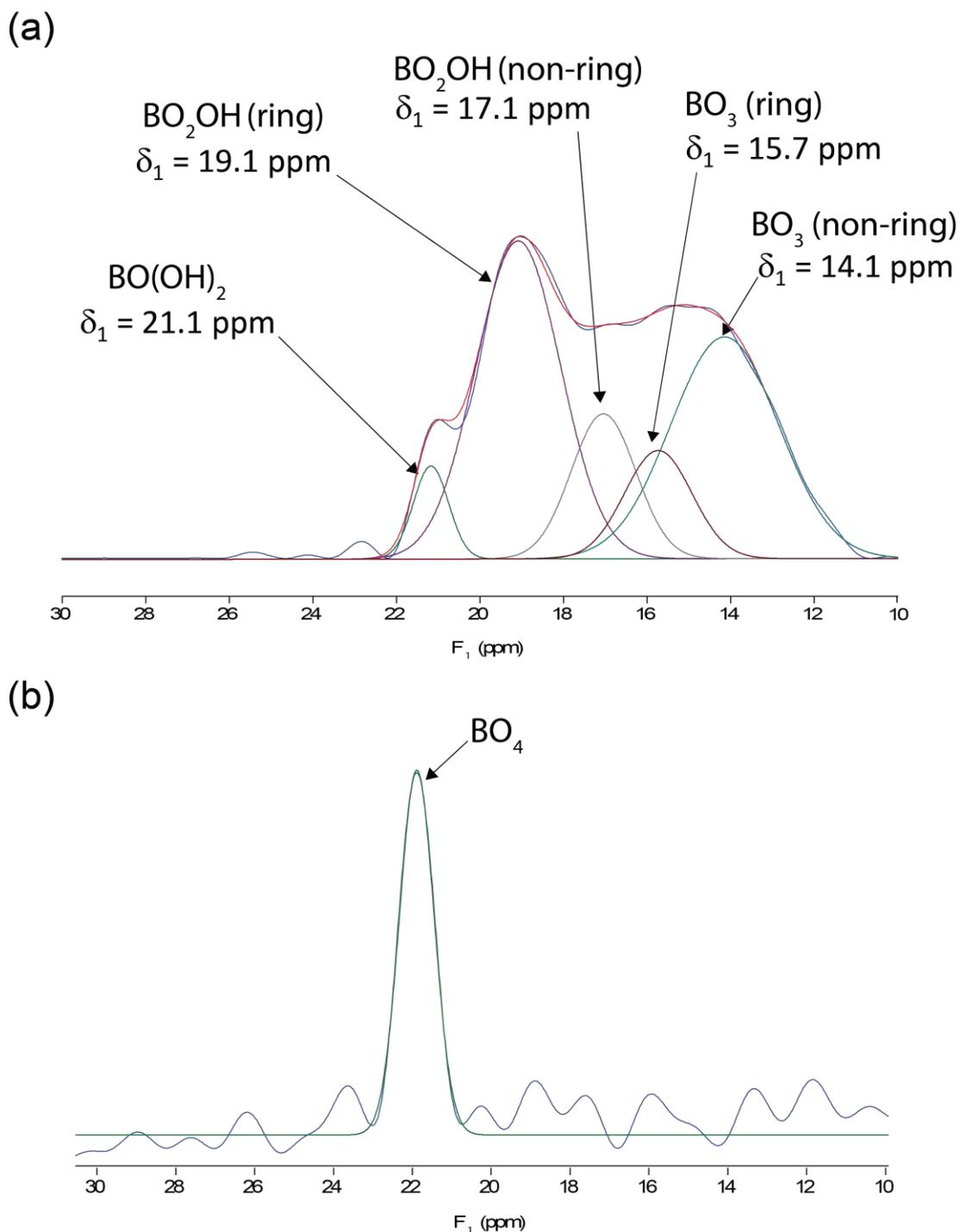

Figure S6. (a-c) Deconvolutions of the lineshapes obtained from total projections of the $F_1$ dimension of the $^{11}$B MQMAS spectrum of DFNS/BN after catalysis. Fitting was carried out using the Gaussian/Lorentzian model in DMFit.[17] The BO$_4$ peak in the subfigure b is folded.

Note: In a sheared MQMAS spectrum, $\delta_1$ is a linear combination of $\delta_2$ and $\delta_{1(\text{non-sheared})}$. It does **not** directly correspond to $\delta_{iso}$ (the isotropic chemical shift). For more information, see the work of Amoureux and Fernandez[8] and Amoureux and Pruski.[18]

**Table S1**. Possible boron species present in DFNS supported catalysts obtained from fitting the $F_1$ dimensions of MQMAS spectra of DFNS/BN before and after catalysis, and DFNS/$B_2O_3$ before catalysis.[19-36] It should be noted that the $\delta_1$ values presented here result from a linear combination of $\delta_2$ and $\delta_{1(non-sheared)}$. They do **not** directly correspond to $\delta_{iso}$ (the isotropic chemical shift).[8,18]

| Sample | Species | $\delta_1$ (ppm) |
|---|---|---|
| As-prepared DFNS/BN | $BN_3$ | 30.5 |
| | $BN_2OH$ | 28.6 |
| | $BN_2O$ | 27.2 |
| | $BO_2OH$(ring) | 18.4 |
| | $BO_2OH$(non-ring) | 17.2 |
| | $BO_3$(ring) | 16.3 |
| | $B(OSi(Q^3))_3OH$ | 0.1 |
| | $B(OB)_4$ | −1.0 |
| | $B(OSi(Q^4))_4$ | −1.9 |
| As-prepared DFNS/$B_2O_3$ | $BO(OH)_2$ | 20.5 |
| | $BO_2OH$(ring) | 19.0 |
| | $BO_2OH$(non-ring) | 17.2 |
| | $BO_3$(ring) | 15.8 |
| | $BO_3$(non-ring) | 14.4 |
| | $BO_4$ | a |
| DFNS/BN after catalysis | $BO(OH)_2$ | 21.1 |
| | $BO_2OH$(ring) | 19.1 |
| | $BO_2OH$(non-ring) | 17.1 |
| | $BO_3$(ring) | 15.7 |
| | $BO_3$(non-ring) | 14.1 |
| | $BO_4$ | a |

aValue not reported because peak was folded.[18,37]

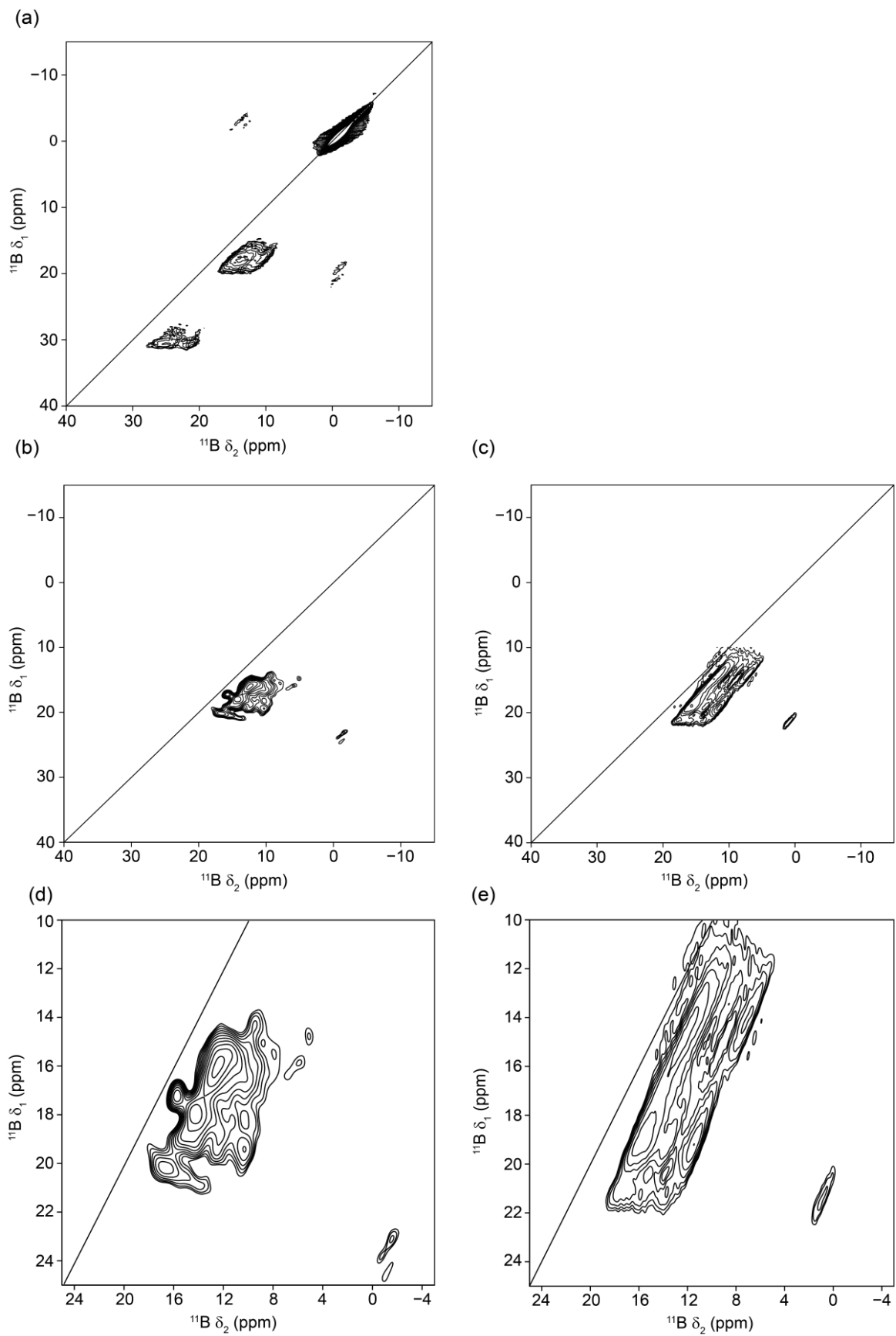

**Figure S7.** 2D $^{11}$B ($^{1}$H decoupled) z-filtered MQMAS spectra of (a) as-prepared DFNS/BN, (b, d) as-prepared DFNS/B$_2$O$_3$ and (c, e) DFNS/BN after catalysis. Subfigures (d) and (e) are expansions of subfigures (b) and (c), respectively. Experiments were recorded at B$_0$ = 18.8 T with ν$_R$ = 20 kHz.

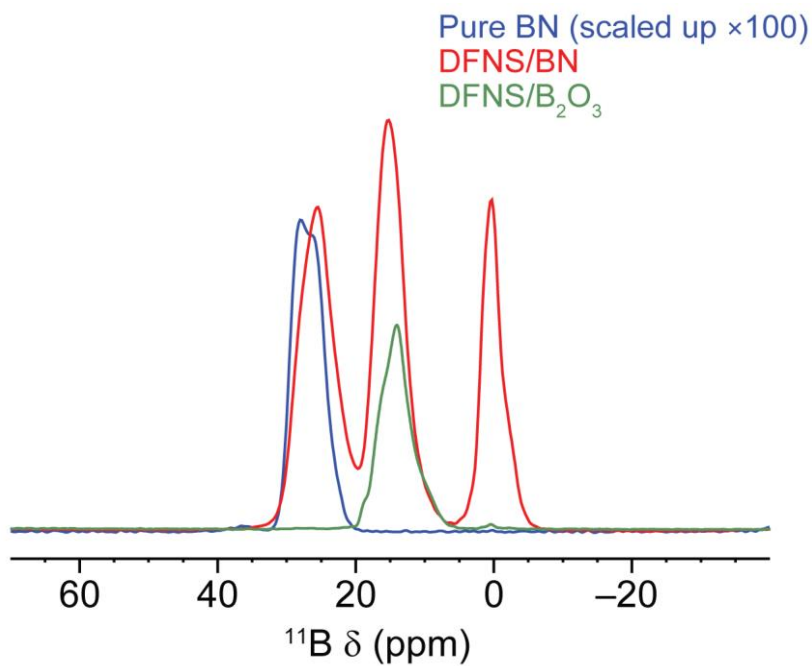

**Figure S8**. $^{11}$B quantitative (*i.e.*, short tip angle) 1D NMR spectra of pure BN, as-prepared DFNS/BN and DFNS/B$_2$O$_3$. Experiments were recorded at B$_0$ = 18.8 T with ν$_R$ = 20 kHz.

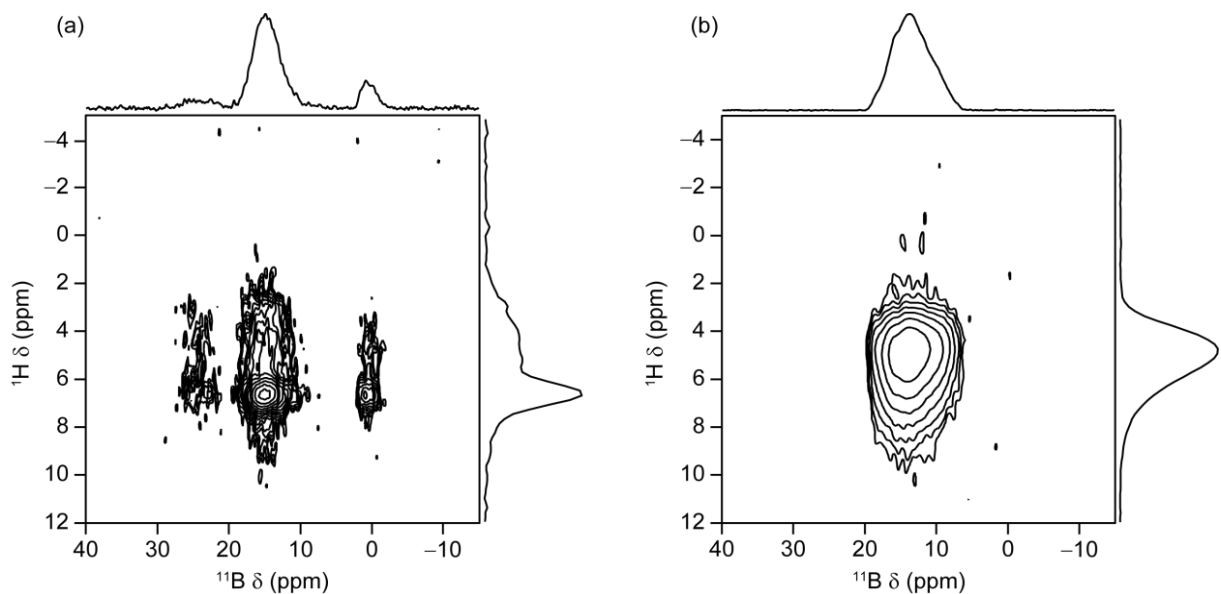

**Figure S9**. Comparison of 2D $^{11}$B-{$^1$H} *D*-HMQC spectra of (a) as-prepared DFNS/BN and (b) DFNS/BN after catalysis.

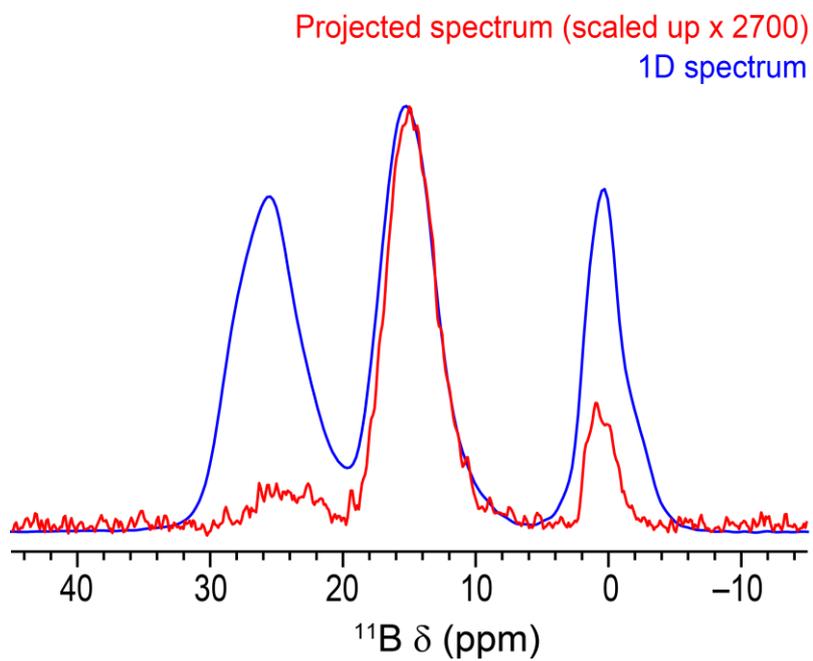

**Figure S10**. Comparison of 1D $^{11}$B quantitative (short tip angle) spectrum and lineshape obtained from the total projection of the F$_2$ dimension of the $^{11}$B-{$^1$H} *D*-HMQC spectrum of as-prepared DFNS/BN.

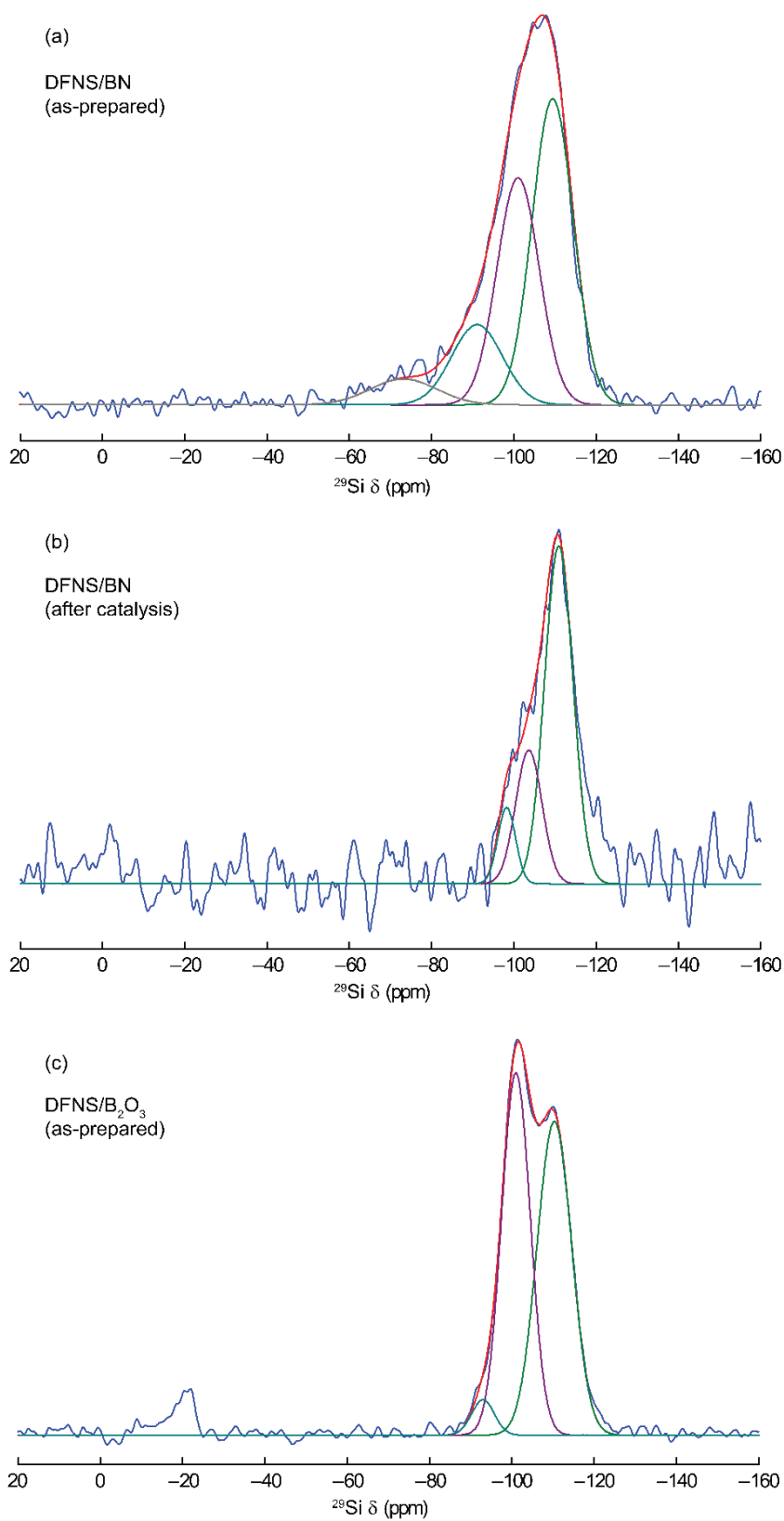

**Figure S11**. Deconvolutions of the lineshapes obtained from $^1H \rightarrow {}^{29}Si$ CP MAS experiments on (a) as-prepared DFNS/BN, (b) DFNS/BN after catalysis and (c) as-prepared DFNS/B$_2$O$_3$. Fitting was carried out using the Gaussian/Lorentzian model in DMFit.[17] In (a), the small peak at *ca.* −20 ppm in the $^1H \rightarrow {}^{29}Si$ CP MAS spectrum of as-prepared DFNS/B$_2$O$_3$ is probably due to residual organic species, so was not included in the fitting process.

**Table S2**. Data obtained from the deconvolutions of the $^1\text{H}\rightarrow{}^{29}\text{Si}$ CP MAS NMR spectra (using the Gaussian/Lorentzian model in DMFit.[17]) presented in Figure S6 above. Parameters include: peak assignment;[38-48] chemical shift, δ; full width at half maximum, FWHM; absolute integrated area (%).

| Sample | Assignment | δ (ppm) | FWHM (ppm) | Area (%) |
|---|---|---|---|---|
| As-prepared DFNS/BN | $SiN_2O_2$ | −73.2 | 19.1 | 6.2 |
|  | $Q^2$ | −91.1 | 14.6 | 14.6 |
|  | $Q^3$ | −101.0 | 12.1 | 34.2 |
|  | $Q^4$ | −109.5 | 11.8 | 45.0 |
| DFNS/BN after catalysis | $Q^2$ | −98.3 | 5.1 | 9.3 |
|  | $Q^3$ | −103.7 | 7.6 | 24.4 |
|  | $Q^4$ | −110.9 | 8.1 | 66.3 |
| As-prepared DFNS/$B_2O_3$ | $Q^2$ | −93.1 | 6.5 | 3.6 |
|  | $Q^3$ | −101.0 | 8.2 | 47.1 |
|  | $Q^4$ | −110.4 | 9.9 | 49.3 |